\documentclass[12pt]{JHEP3}
\usepackage{graphicx,amsfonts,amsmath,times,cite}

\preprint{IPhT-T09/041}
\title{Boundary changing operators in the $O(n)$ matrix model}
\author{Jean-Emile Bourgine$^1$\\ \\
$^1$ Institut de Physique Th\'eorique, CNRS-URA 2306  \\ ~~
     C.E.A.-Saclay, F-91191 Gif-sur-Yvette, France    \\
\email{jean-emile.bourgine@cea.fr}
}

\abstract{
  We continue the study of boundary operators in the dense $O(n)$ model on the random lattice. The conformal dimension of boundary operators inserted between two JS boundaries of different weight is derived from the matrix model description. Our results are in agreement with the regular lattice findings. A connection is made between the loop equations in the continuum limit and the shift relations of boundary Liouville 3-points functions obtained from Boundary Ground Ring approach.
}

\begin{document}
\newcommand{\tr}{{\rm tr~}}
\newcommand{\Tr}{{\rm Tr~}}
\newcommand{\FIG}[4]{\FIGURE[t]{\includegraphics[width=#1cm]{#2}\caption{#3}\label{#4}}}

  %%%%%%%%%GREEK LETTERS%%%%%%%%%%%%%%%%
 \def\a{\alpha}
 \def\b{\beta}
 \def\g{\gamma}
 \def\d{\delta}
 \def\e{\epsilon}
 \def\te{\tilde{E}}
 \def\eps{\varepsilon}
 \def\th{\theta}
 \def\vt{\vartheta}
 \def\k{\kappa}
 \def\l{\lambda}
 \def\m{\mu}
 \def\n{\nu}
 \def\x{\xi}
 \def\r{\rho}
 \def\vr{\varrho}
 \def\s{\sigma}
 \def\t{\tau}
 \def\th{\theta}
 \def\z{\zeta }
 \def\vp{\varphi}
 \def\G{\Gamma}
 \def\D{\Delta}
 \def\T{\Theta}
 \def\X{\Xi}
 \def\P{\Pi}
 \def\S{\Sigma}
 \def\L{\Lambda}
 \def\O{\Omega}
 \def\oo{\hat \omega   }
 \def\ov{\over}
 \def\o{\omega }
\def\p{\partial}
%----My defs :
\def\Disc{\text{Disc\ }}
\def\la{\left\langle}
\def\ra{\right\rangle}
\def\bL{b_\phi}
\def\pa{^{(1\parallel)}}
\def\pe{^{(0\perp)}}
\def\paa{^{(0\parallel)}}
\def\pij{^{(1 I\cap J)}}
\def\Card{\text{Card}\ }
\def\ee{^{(E)}}
\def\le2{^{(L\ E_2)}}
\def\le2{^{(L\ 1\mid E_1\ E_2)}}

%----------------------------Corps du texte
\section{Introduction}
Boundary conformal field theories  have led to a wide range of applications in theoretical physics. From the study of fixed point in quantum impurity problems to the description of branes in open string theory, conformally invariant boundary conditions are the key point of many fascinating problems. Unfortunately several aspects still need a better comprehension, such as the classification of boundary conditions or the fusion of boundary operators in non-rational conformal field theories (CFT).

In this article we concentrate on the $O(n)$ model which provides a microscopic description of (non-rational) conformal field theories with central charge $c<1$. The $O(n)$ model can also be investigated on a fluctuating lattice and the introduction of the matrix model proved to be a powerful tool for the analysis of both bulk \cite{Kostov:1991cg,Kazakov:1986hy,Boulatov:1986sb,Duplantier:1988wc,Kazakov:1988fv,Kostov:1988fy} and boundary \cite{Kazakov:1991pt,Kostov:2003uh,Kostov:2002uq} behavior. In the continuum limit, the CFT is coupled to Liouville gravity and can be interpreted as a string theory with target space dimension smaller than two\cite{Kostov:1991cg}.

The $O(n)$ model is considered on a triangular lattice $\G$, to each face is associated a classical $O(n)$ spin $S_a(r),a=1\cdots n$. The partition function describes nearest neighbor interactions, the coupling constant being given by the inverse temperature $T^{-1}$ \cite{Nienhuis:1982fx},\cite{Nienhuis:1984wm},
\begin{equation}
Z_\G(T)=\tr{\prod_{<rr'>}{\left(1-\dfrac{1}{T}\sum_a{S_a(r)S_a(r')}\right)}}.
\end{equation}
Paths of identical spins draw self and mutually avoiding loops of weight $n$ on the lattice. The partition function can be re-expressed as a sum over loop patterns \cite{Kostov:1991cg},
\begin{equation}
Z_\Gamma(T)=\sum_{\text{loops}}T^{-(\text{length})}n^{\#(\text{loops})},
\end{equation}
where the temperature is coupled to the total length of the loops. This formulation allows analytic continuation of the parameter $n$ to arbitrary real value. For fixed values of $-2\leq n\leq 2$ the model develops two phase transitions with respect to the temperature \cite{Kostov:1991cg,Kostov:2006ry}. We will only be interested in the dense phase, and consider the model at zero temperature where loops are fully packed. When put on a dynamical lattice, this point renormalizes to a CFT \cite{DiFrancesco:1999} with central charge 
\begin{equation}
c_\text{dense}=1-\dfrac{6\th^2}{1-\th},\quad n=2\cos{\pi\th},\quad \th\in[0,1]
\end{equation}
For the critical values $n=2\cos{\dfrac{\pi}{h}}$, $h\in\mathbb{Z}$ this CFT is essentially the rational $(h,h-1)$ minimal model.

Following the work of \cite{Nichols:2004fb,Pearce:2006sz}, a continuous set of boundary conditions was discovered by Jacobsen and Saleur \cite{Jacobsen:2006bn,Jacobsen:2008}. Such boundary conditions, referred as JS boundary conditions, are introduced as follows. On the boundary a certain subset of the spin components is allowed to fluctuate. Fixing $n-k$ components, the $O(n)$ symmetry is broken into $O(k)\times O(n-k)$ but the conformal invariance is still present at the critical point. When reformulated in the loop gas language, this boundary condition gives a weight $k$ to loops that touch the boundary. The parameter $k$ can be analytically continued to arbitrary real values and interpolates between Dirichlet ($k=1$) and Neumann ($k=n$) boundary conditions. This continuous set of boundary conditions is conveniently parameterized by
\begin{equation}\label{paramJS}
k(r)=\dfrac{\sin{\pi (r+1)\th}}{\sin{\pi r\th}},\quad r\in\left[0,\frac{1}{\th}\right].
\end{equation}

Additionally $L$ open lines starting or ending between two boundaries are introduced. These ``$L$-leg'' boundary operators can be seen as the fusion of the boundary changing operator with the star operator of \cite{Kostov:2003uh}. Two different sectors must be discussed, named blobbed and unblobbed, this term referring to the underlying Temperley-Lieb algebra. In the blobbed sector, the open line next to the JS boundary is allowed to touch it, whereas this is forbidden in the unblobbed sector. Reformulated in the spin language, the spin path described by a blobbed (resp. unblobbed) open line has components corresponding to the fluctuating (resp. fixed) boundary spin.

In \cite{Jacobsen:2006bn}, the $O(n)$ model was considered on an annulus with Neumann and JS boundary conditions on respectively the inner and outer rims. Furthermore, $L$ non-contractible blobbed/unblobbed loops surrounding the inner rim were also introduced. The partition function was computed, leading to a conjecture for the scaling dimension of boundary changing operators. This conjecture was further checked on the random lattice in \cite{Kostov:2007jj} where a disc with Neumann and JS boundaries was studied. This disc partition function can be generated as a matrix model correlator and the previous results were re-derived in this context in \cite{BH:2009}.

Boundary changing operators between two JS boundaries with different parameter $k$ were discussed in \cite{Dubail:2008} and the corresponding annulus partition function computed. The purpose of the present article is to carry out a similar investigation on the fluctuating lattice in order to compare the spectrum of JS-JS boundary operators with the results derived on the flat lattice. In a first section we will recall the boundary $O(n)$ matrix model and some useful results from Liouville theory. Then, both sectors with and without open lines will be considered in respectively section 2 and 3. Scaling exponents will be recovered in the continuum limit and contact with the boundary Liouville 3-points function will be made. Technical points that can be omitted at the first reading are gathered in the Appendix.
\vspace{1cm}

\textbf{Summary of the results:}
We consider the $O(n)$ model on a random lattice with the disc topology, and impose to the boundary spins one Neumann and two JS boundary conditions of parameter $k_I$ and $k_J$. The insertion of open lines between the boundaries modify the spectrum of the JS-JS boundary operator, and we need to differentiate two sectors. In both sectors disc partition functions can be obtained as matrix model correlators. Using the standard matrix model technique, we derived a set of loop equations and analyzed their continuum limit. In this way, we found the gravitational scaling of the correlators and, via the KPZ relation, recovered the critical dimension of the JS-JS boundary operators. As a consistency check the critical loop equations were mapped on boundary ground ring relations obeyed by Liouville correlators.

In the closed loop sector, no open lines are introduced between the two JS boundaries and loops touching both boundaries can form. Thus, the scaling dimension of the JS-JS boundary operators also depends on the weight $k_{IJ}$ assigned to those loops. Parameterizing the weight of the loops as \ref{paramJS} and \cite{Dubail:2008}
\begin{equation}\label{paramDJS}
k_{IJ}(r_{IJ})=\dfrac{\sin{(r_I+r_J+1-r_{IJ})\pi\th/2}\ \sin{(r_I+r_J+1+r_{IJ})\pi\th/2}}{\sin{r_I\pi\th}\ \sin{r_J\pi\th}},
\end{equation}
with $r_{IJ}\in [1,1+2/\th]$, the scaling dimension of the JS-JS boundary operators belongs to
\begin{equation}\label{res1}
\d_{r_{IJ}+2j,r_{IJ}},\quad j\in \mathbb{Z},
\end{equation}
where we used the Kac notation \ref{Kac}.

When $L$ open lines are inserted, loops touching both boundaries are forbidden and the scaling dimension of the JS-JS boundary operators depends only on $k_I$, $k_J$ and $L$,
\begin{equation}\label{res2}
\d_{\e_Ir_I+\e_Jr_J+1+2j,\e_Ir_I+\e_Jr_J+1-L},\quad j\in \mathbb{Z}^+,
\end{equation}
where the sign $\e_I=\pm$ (resp. $\e_J$) is plus when the open lines are blobbed with respect to the JS boundary of parameter $k_I$ (resp. $k_J$).

\section{Preliminaries}
\subsection{The $O(n)$ matrix model with boundaries}
The $O(n)$ partition function on a fluctuating lattice is obtained by summing the regular lattice partition functions over realizations of a random lattice. In the case of the disc topology it writes,
\begin{equation}
 Z_{\text{dyn}}(\kappa,x,T)=
 \sum_{\Gamma:\;\text{disc}}\frac1{L(\Gamma)}
 \kappa^{-A(\Gamma)}x^{-L(\Gamma)}Z_\Gamma(T),
\end{equation}
where two cosmological constants were introduced. The bulk cosmological constant $\k$ controls the area $A(\G)$ of lattices and the boundary cosmological constant $x$ the length $L(\G)$ of the boundary. Such a boundary cosmological constant must be introduced for each different boundary. In the following, we will use the convention to denote respectively by $x$ and $y$ the boundary cosmological constants of Neumann and JS boundaries type.

Disc partition functions on a random lattice are obtained as the planar limit of the $O(n)$ matrix model simple trace correlators \cite{BH:2009}. At the zero temperature point, the partition function of the $O(n)$ matrix model \cite{Kostov:1988fy} reduces to
\begin{equation}
Z=\int{dX\prod_{a=1}^n{dY_a}\exp{\left[-\b\tr{\left(\dfrac{1}{2}X^2+\dfrac{1}{2}\sum_{a=1}^n{Y_a^2}-\sum_{a=1}^n{XY_a^2}\right)}\right]}},
\end{equation}
$X$ and $Y_a$ being $N\times N$ hermitian matrices. The planar limit is achieved by sending the size of the matrices $N$ and the parameter $\b$ to infinity, keeping the cosmological constant $\k$ finite,
\begin{equation}
\b=N\k^2.
\end{equation}

The disc partition function with Neumann boundary condition is given by the first order in the large $N$ limit of the correlator\footnote{It is also possible to obtain Neumann boundary conditions as JS boundary conditions with $k=n$. On the random lattice these two Neumann partition functions are inverse of eachother because the two boundaries have different dimension.}
\begin{equation}
\Phi(x)=-\dfrac{1}{\b}\la\tr{\log{(x-X)}}\ra.
\end{equation}
Here the quantity of importance is actually its derivative, the resolvant
\begin{equation}
W(x)=-\dfrac{\p}{\p x}\Phi(x)=\dfrac{1}{\b}\la\tr{\dfrac{1}{x-X}}\ra
\end{equation}
where one point on the boundary has been marked. This quantity plays a special role in the study of matrix models. In the planar limit, it is known to have a branch cut on the support of the eigenvalue density $[a,b]\subset\mathbb{R}^-$. This branch cut is a common property of correlators involving Neumann boundaries.

To study the disc partition function with two JS boundaries, we need to introduce a third boundary of Neumann kind because loop equations couple both JS and Neumann boundaries. To each JS boundary we associate an integer subset $I\subset [1,n]$ with $k_I$ elements corresponding to the spin components allowed to fluctuate. The matrix operator that creates the JS$_I$ boundary with cosmological constant $y_I$ will be denoted
\begin{equation}
H_I=\dfrac{1}{y_I-\sum_{a\in I}{Y_a^2}}.
\end{equation}

The matrix operators that introduce the open lines were defined in \cite{BH:2009}. Here we need to slightly generalize their definition because open lines introduced between JS boundaries can be blobbed with respect to zero, one or both boundaries. Let $E$ be an integer subset of $[1,n]$, we define the $L$-legs matrix operators as
\begin{equation}
Y_L^{(E)}=\sum_{\{a_1,\cdots,a_L\}\subset E}{Y_{a_1}\cdots Y_{a_L}},
\end{equation}
where the sum is taken over all sets of unequal indices $a_i\neq a_j$. If $E\subset I$ then the operator is said to be blobbed with respect to the JS$_I$ boundary. On the contrary, when $E\cap I=\emptyset$, the operator is unblobbed with respect to this boundary. These open lines must be counted with a weight $k_E=\Card E$.

In \cite{BH:2009} were introduced matrix correlators corresponding to disc with mixed Neumann-JS boundaries and $L$ open lines between them,
\begin{equation}
D_I^{(L\ E)}(x,y_I)=\dfrac{1}{\b}\la\tr{\dfrac{1}{x-X}Y_L^{(E)}H_IY_L^{(E)\dagger}}\ra.
\end{equation}
They were denoted $D_I^{(L\ \parallel)}$ when $E=I$, i.e. blobbed open lines with weight $k_I$, and $D_I^{(L\ \perp)}$ when $E=[1,n]\setminus I$ corresponding to unblobbed open lines with weight $n-k_I$.

We now extend the previous definitions to the three boundaries Neumann-JS$_I$-JS$_J$ case, where $I$ and $J$ are two integer subset of $[1,n]$ of cardinal respectively $k_I$ and $k_J$. Their intersection describes loops touching both boundaries, such loops having a weight $k_{IJ}=\Card I\cap J$. The disc correlator without open lines is given by
\begin{equation}
D_{IJ}\pe(x,y_I,y_J)=\dfrac{1}{\b}\la\tr{\dfrac{1}{x-X}H_IH_J}\ra.
\end{equation}

Considering two integer subsets $E_1$ and $E_2$ of $[1,n]$ let us introduce the disc correlator with an  insertion of $L_1$ open lines between the boundaries Neumann-JS$_I$ and JS$_I$-JS$_J$, and $L_2$ open lines between the boundaries Neumann-JS$_J$ and JS$_I$-JS$_J$,
\begin{equation}
D_{IJ}^{(L_1\ L_2\mid E_1\ E_2)}(x,y_I,y_J)=\dfrac{1}{\b}\la\tr{\dfrac{1}{x-X}Y_{L_1}^{(E_1)}H_IY_{L_1}^{(E_1)\dagger}Y_{L_2}^{(E_2)}H_JY_{L_2}^{(E_2)\dagger}}\ra.
\end{equation}
If we consider any $E_1$ and $E_2$, there is a redundancy coming from open lines starting between Neumann-JS$_I$ boundaries, bouncing at the frontier JS$_I$-JS$_J$ and ending between Neumann and JS$_J$ boundaries. To forbid such an open line bouncing at some point, we need to impose $E_1\cap E_2=\emptyset$. All these quantities satisfy the reflection property
\begin{equation}\label{reflection}
D_{IJ}(x,y_I,y_J)=D_{JI}(x,y_J,y_I).
\end{equation}

It will be more convenient to use a shortcut notation when only one set of open lines is involved,
\begin{align}
\begin{split}\label{maindef}
&D_{\bar{I}J}^{(L\ E)}(x,y_I,y_J)=\dfrac{1}{\b}\la\tr{\dfrac{1}{x-X}Y_L^{(E)}H_IY_L^{(E)\dagger}H_J}\ra,\\
&D_{I\bar{J}}^{(L\ E)}(x,y_I,y_J)=\dfrac{1}{\b}\la\tr{\dfrac{1}{x-X}H_IY_L^{(E)}H_JY_L^{(E)\dagger}}\ra,\\
&D_{\bar{I}J}^{(E)}(x,y_I,y_J)=\dfrac{1}{\b}\la\tr{\dfrac{1}{x-X}Y_1^{(E)}H_IY_1^{(E)\dagger}H_J}\ra,\\
&D_{I\bar{J}}^{(E)}(x,y_I,y_J)=\dfrac{1}{\b}\la\tr{\dfrac{1}{x-X}H_IY_1^{(E)}H_JY_1^{(E)\dagger}}\ra.
\end{split}
\end{align}

We can also insert $L$ open lines between Neumann-JS$_I$ and Neumann-JS$_J$ boundary points,
\begin{align}
\begin{split}\label{DIJL}
&D_{IJ}^{(L\parallel)}(x,y_I,y_J)=\dfrac{1}{\b}\la\tr{\dfrac{1}{x-X}Y_L^{(I\cap J)}H_IH_JY_L^{(I\cap J)\dagger}}\ra,\\
&D_{IJ}^{(L\perp)}(x,y_I,y_J)=\dfrac{1}{\b}\la\tr{\dfrac{1}{x-X}Y_L^{(\overline{I\cup J})}H_IH_JY_L^{(\overline{I\cup J})\dagger}}\ra.
\end{split}
\end{align}
where $\overline{I\cup J}$ designate the complementary set of $I\cup J$ in $[0,n]$.

In the continuum limit all parameters $k_I$, $k_J$, $k_{IJ}$ will be analytically continued to any real value. For clarity reasons the dependence of correlators in the JS boundary cosmological constants will be hidden whenever this dependence is not directly relevant.

\subsection{Continuum limit and boundary Liouville correlators}
In this section we recall briefly the main ideas of the continuum limit for the matrix model correlators, more details on this subject can be found in \cite{Kostov:2003uh,Kostov:2002uq,BH:2009}. We also present some useful results of boundary Liouville theory, we address the reader to the original papers \cite{Fateev:2000ik, Ponsot:2001ng, Hosomichi:2001xc} for further interest.

The continuum limit is achieved by sending the cosmological constants to their critical values where the average area and the average length of boundaries diverge. The renormalized coupling constants are defined by blowing up the region near the critical point,
\begin{equation}\label{contlim}
\e^{2g}\m=\k-\k^*,\quad \e\xi=x-x^*,\quad \e^g\z=y-y^*,
\end{equation}
where was introduced the elementary length $\e^g$ of the lattice as a cut-off. Boundaries with Neumann boundary conditions have a fractal dimension $1/g$, $g=1-\th$ so that their boundary cosmological constant scales as $\xi\sim\m^{1/2g}$. Boundaries of JS kind have the usual dimension one and the cosmological constant simply scales as $\z\sim\m^{1/2}$. The critical value for the boundary cosmological constant were determined in the previous studies \cite{Kostov:1991cg}, \cite{BH:2009}, $x^*=0$ and $y^*=(k+1)W(0)$.

Let us consider a disc correlator with one Neumann boundary and an arbitrary number of JS boundaries $D(x,y_i)$. The continuum limit of this correlator is obtained by subtracting the non-critical part,
\begin{equation}
\e^{\a}d(\xi,\z_i)=D(x,y_i)-D^*(x,y_i).
\end{equation}
This non-critical part $D^*$ represents special limits of the disc correlator where one or more boundary disappear, it must vanish after a finite number of derivative with respect to the boundary cosmological constants.\footnote{This interpretation of the non-critical term of matrix model correlators is further explicited in \cite{Alexandrov:2005}.}. The critical correlator $d$ as function of $\xi$ has a branch cut on $]-\infty,-M]\subset\mathbb{R}^-$, where $M$ is a function of the bulk cosmological constant and was computed in \cite{Kostov:2006ry}.

Critical points of statistical models on a random lattice are described in the continuum limit as a CFT coupled to 2D gravity. In Polyakov gauge the effective degree of freedom for the gravity is a Liouville field $\phi$. The Liouville theory is conformal and coupling to the conformal matter part is achieved through the requirement of vanishing total central charge
\begin{equation}
c_\text{tot}=c_\text{CFT}+c_\text{Liouville}+c_\text{ghost}.
\end{equation}
This coupling gives the value $b=\sqrt{g}$ to the Liouville parameter. Operators are dressed by ghost and Liouville fields and we have to sum over the position of insertion in order to get diffeomorphism invariant quantities. We also require the vanishing of the total scaling dimension
\begin{equation}
\D_\text{matter}+\D_\text{Liouville}+\D_\text{ghost}=0.
\end{equation}
This requirement induces an important relation between the bare scaling dimension of CFT operators $\d$ and the Liouville momentum of the dressing factors $P$. We will extensively make use of the Kac notation for these two quantities,
\begin{equation}\label{Kac}
\d_{r,s}=\dfrac{(r/b-sb)^2-(1/b-b)^2}{4},\quad P_{r,s}=\dfrac{r}{2b}-\dfrac{sb}{2},
\end{equation}
but allow non integer indices $(r,s)$. Correlators of 2D gravity factorize into matter, ghost and Liouville parts. The study of the Liouville part is sufficient to determine the scaling exponents of bare operators via the dressing momenta. The Liouville term is also the only dependence on the boundary cosmological constants, through the boundary parameters $(\t,\s)$ specified as (\cite{BH:2009})
\begin{equation}\label{param}
\xi(\t)=M\cosh{\t},\quad \z(\s)=\dfrac{M^g}{2g}\dfrac{\sin{\pi\th}}{\sin{\pi r_I\th}}\cosh{g\s}.
\end{equation}
Consequently, we can neglect the ghost and matter part, focusing on the boundary parameters dependence. Let us define the Liouville boundary operator of momentum $P$ inserted between two boundaries labeled by $\t$ and $\s$, as
\begin{equation}
^{\t}B_P^{\ \s}=e^{(Q/2-P)\phi}.
\end{equation}
with the Liouville charge $Q=b+\frac{1}{b}$. Here our convention slightly differs from the one used in \cite{BH:2009}. Most of the time, the momentum $P$ will be positive and this definition corresponds to the usual dressing of boundary CFT operators in the matrix model \cite{Kostov:2003uh}. It may happen that some momenta reveal to be negative, then the operator is found to have a wrong dressing  with respect to 2D gravity. All Liouville boundary operators obey the reflection relation
\begin{equation}\label{reflectBO}
^{\t}B_{P}^{\ \s}=4P\ d(P\mid\t,\s)\ \cdot\ ^{\t}B_{-P}^{\ \s}
\end{equation}
involving the Liouville boundary 2-points function,
\begin{equation}
d(P\mid\s,\t)=\la ^{\t}B_{P}^{\ \ \s}B_{P}^{\ \ \t}\ra.
\end{equation}
We will not need the explicit expression in terms of double sine functions found in \cite{Fateev:2000ik} (see also \cite{Kharchev:2001rs}), but the shift relation
\begin{equation}\label{shift2pt}
\dfrac{\sin{\pi\p_\t}}{C\sinh{b^2\t}}d(P\mid\t,\s)=d(P-b/2\mid\t,\s),
\end{equation}
where $C$ is some constant independent of the boundary parameters, and the reflection property\footnote{An irrelevant factor of $8P^2$ was intentionnaly omitted for a matter of simplicity.}
\begin{equation}\label{reflect2pt}
d(-P\mid\t,\s)=d(P\mid\t,\s)^{-1}
\end{equation}
will be useful.

The gravitational scaling of the Liouville correlator $\g$ is linked to the momenta of boundary operators via the KPZ formula\cite{Knizhnik:1988ak,David:1988hj,Distler:1988jt},
\begin{equation}
 \big\langle {}^{\s_1}\!B_{P_1}\!\!{}^{\s_2}\cdots
            {}^{\s_n}\!B_{P_n}\!\!{}^{\s_1}\big\rangle \propto \mu^\gamma,
\end{equation}
with
\begin{equation}\label{KPZ}
 2b\gamma=\big(1-\frac n2\big)\big(b+\frac{1}{b}\big)+\sum_{i=1}^n{P_i}.
\end{equation}

The disc with mixed Neumann-JS boundaries was studied by matrix model technique in \cite{BH:2009}. Solving the loop equation in the continuum limit, the boundary Liouville 2-points function was recovered. The Liouville momentum of the dressed operator changing from Neumann to JS boundary condition of parameter $k_I(r_I)$ and with $L$ open lines inserted was found to be
\begin{equation}\label{momenta2pt}
P_I^{(L\perp,\parallel)}=\pm r_I\left(\dfrac{1}{2b}-\dfrac{b}{2}\right)+L\dfrac{b}{2}
\end{equation}
where the plus sign stands for unblobbed open lines. Such a momentum assign a gravitational scaling $\g_I=\frac{P_I}{2b}$ to matrix correlators. All momenta excepted $P\paa$ and $P\pa$ are positive. The operator carrying momentum $P\paa$ will be of no use here because the boundary operator with no open line inserted has always momentum $P\pe$. Furthermore, if we restrict to $k_I>0$ (i.e. $r_I<(1-\th)/\th$), $P\pa$ is positive. When $P\pa$ is negative, a wrong dressing of the bare operator has to be used.

Let us now focus on the three boundaries case and denote by $d_{IJ}$ the continuum limit of the correlator
\begin{equation}\label{contlim3pts}
\e^{\a_{IJ}}d_{IJ}(\xi,\z_I,\z_J)=D_{IJ}(x,y_I,y_J)-D_{IJ}^*(x,y_I,y_J)
\end{equation}
where $\a_{IJ}$ relates to the gravitational scaling $\m^{\g_{IJ}}$ by $\a_{IJ}=2b^2\g_{IJ}$. The expression of the critical part $d_{IJ}$ is given by the Liouville boundary 3-points function\footnote{up to a factor independent of the boundary parameters, corresponding the ghost and matter part of the correlator.}
\begin{equation}
d(P_I,P,P_J\mid \s_I,\s_J,\t)=\la ^\t B_{P_I}^{\ \s_I} B_{P}^{\ \s_{J}} B_{P_J}^{\ \t}\ra.
\end{equation}
The exact expression of this function was found by Ponsot and Teschner in \cite{Ponsot:2001ng}. Here we will only need some properties such as the reflection and cyclic symmetries,
\begin{align}
\begin{split}
&d(P_1,P_2,P_3\mid\s_1,\s_2,\s_3)=d(P_3,P_2,P_1\mid\s_2,\s_1,\s_3),\\
&d(P_1,P_2,P_3\mid\s_1,\s_2,\s_3)=d(P_2,P_3,P_1\mid\s_2,\s_3,\s_1)=d(P_3,P_1,P_2\mid\s_3,\s_1,\s_2).
\end{split}
\end{align}
This function is also known to obey an important shift relation involving both momenta and boundary parameters \cite{BHKM:2007}. This relation simplifies when the momenta obey a specific identity, as explained in Appendix A.1.

Specializing the KPZ formula to the 3-points case, we get a relation that allows us to recover the scaling dimension of the operator inserted between the JS boundaries directly from the gravitational scaling $\g_{IJ}$ of the matrix model correlator. Indeed, the expression for the momenta of the operators inserted between Neumann and JS boundaries is already known to be given by \ref{momenta2pt}. Then we directly extract
\begin{equation}\label{KPZ3pts}
P=2b\g_{IJ}+\dfrac{1}{2b}+\dfrac{b}{2}-P_I-P_J.
\end{equation}

As an example, let us consider the simplest case of $d_{IJ}\pe$ correlator. No open lines are inserted between Neumann and JS boundaries so that the momenta of the boundary operators are simply given by $P_I\pe$, $P_J\pe$ of \ref{momenta2pt}. Using \ref{KPZ3pts} we deduce from the gravitational dimension $\g_{IJ}\pe$ the  momentum of the operator changing JS$_I$ into JS$_J$ boundary condition
\begin{equation}
P_{IJ}\pe = 2b\g_{IJ}\pe-\dfrac{r_I+r_J-1}{2b}+(r_I+r_J+1)\dfrac{b}{2}.
\end{equation}
Anticipating over the next section, let us mention that this correlator $d_{IJ}\pe$ is coupled to $d_{IJ}\pa$ in loop equations. The correlator $d_{IJ}\pa$ corresponds to a disc with Neumann-JS$_I$-JS$_J$ boundaries and one open line inserted, starting between Neumann and JS $_I$ boundaries and ending between Neumann and JS$_J$ boundaries. The open line is allowed to touch both JS boundaries and is introduced by boundary operators of momenta $P_I\pa$ and $P_J\pa$. The momentum of the operator changing JS$_I$ to JS$_J$ boundary condition remains the same because no open line is inserted at this point. This translates into the relation
\begin{equation}\label{relscalings}
\g_{IJ}\pa-\g_{IJ}\pe=\g_J\pa-\g_I\pe=\g_I\pa-\g_J\pe
\end{equation}
for the gravitational scalings of $d_{IJ}\pe$ and $d_{IJ}\pa$ correlators.

In the next sections, the gravitational scalings will be read of the matrix model loop equations in the continuum limit. By the use of the previous statements, we will be able to determine the scaling dimensions of the JS$_I$-JS$_J$ boundary changing operator.

\subsection{Derivation of the loop equations}
Using the invariance of the matrix measure, in \cite{BH:2009} was derived a powerful loop equation that can be applied to various quantities $G$,
\begin{equation}\label{le}
\la\tr{GY_a}\ra=\sum_{ij}{\dfrac{1}{\b}\la\dfrac{\p}{\p Y_{aij}}\oint{\dfrac{dx'}{2i\pi}\left(\dfrac{1}{x'-X}G\dfrac{1}{-x'-X}\right)_{ij}}\ra}.
\end{equation}
The contour of integration circles the support of the eigenvalue density that corresponds to the branch cut $]-\infty,-M]$ of the correlators. This loop equation can be interpreted as follows. In the LHS is inserted on the boundary a matrix $Y_a$ that describes the starting point of an open line. Because of the gaussian measure for $Y_a$ matrices, this matrix couple to another $Y_a$ inside $G$ corresponding to the end point of the open line. This operation corresponds to remove an open line from the disc correlator, or loop if the end point is situated nearby the starting point. The action of the $Y_a$ derivative in the RHS is to split the matrix product in two traces. Using the factorization property in the planar limit
\begin{equation}
\la\tr{A}\tr{B}\ra\sim\la\tr{A}\ra\la\tr{B}\ra
\end{equation}
we obtain a convolution of correlators in the length space that translates into a ``star product'' in the boundary cosmological constant space,
\begin{equation}
\left(A\ast B\right)(x)=\oint{\dfrac{dx'}{2i\pi}\dfrac{A(x')-A(x)}{x-x'}B(-x')}.
\end{equation}
This product corresponds to the splitting of the disc into two parts along the open line removed in the LHS. Applied to various disc quantities, the relation \ref{le} provides us with several loop equations that can be used to determine the gravitational dimension of the matrix model correlator.

\section{Closed loops sector}
In this section we will be intersted in finding the scaling dimension of the operator changing boundary conditions JS$_I$ to JS$_J$ but with no open lines inserted between the two boundaries (see Figure \ref{2bL}). Open lines can be introduced between Neumann and JS boundaries without changing the dimension of this operator. Indeed, considering the quantities defined in \ref{DIJL} we can obtain the following equations when $L>0$ by removing one of the open lines,
\begin{align}
\begin{split}\label{dl}
&D_{IJ}^{(L+1 \parallel)}=(k_{IJ}-L)\ W\ast D_{IJ}^{(L\parallel)},\\
&D_{IJ}^{(L+1 \perp)}=(n-k_I-k_J+k_{IJ}-L)\ W\ast D_{IJ}^{(L\perp)}.
\end{split}
\end{align}
We deduce a recursion relation for the gravitational scalings $\g_{IJ}^{(L+1)}=\g_{IJ}^{(L)}+1/2$ in agreement with \ref{KPZ3pts} where the momenta of operators inserted between Neumann and JS boundaries are given by \ref{momenta2pt}, the operator inserted between the two JS boundaries remaining unchanged.

\FIG{4.5}{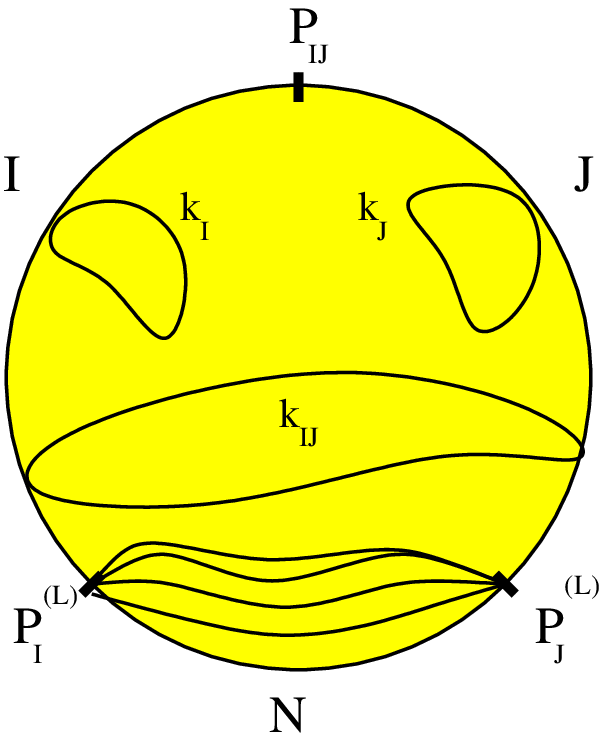}{A disc with no open lines inserted between the two JS boundaries.}{2bL}

\subsection{Coupled loop equations for $D_{IJ}\pe$}
Let us apply the loop equation \ref{le} to $G=\dfrac{1}{x-X}H_IH_JY_c$ and use 
\begin{equation}
\sum_{c\in J}{\la\tr{\dfrac{1}{x-X}H_IH_JY_c^2}\ra}=y_JD_{IJ}\pe-D_I\pe.
\end{equation}
This provides us with a first equation where $D_{IJ}\pe$ and $D_{IJ}\pa$ are coupled,
\begin{equation}\label{equ1}
y_JD_{IJ}\pe=D_I\pe+k_JD_{IJ}\pe\ast W + D_{IJ}\pe\ast D_J\pa+D_I\pe\ast D_{IJ}\pa.
\end{equation}
Each term of the RHS has an interpretation as splitting the LHS correlator in two parts. The first term corresponds to the vanishing of JS$_J$ boundary. The second one reprents the removal of a close loop. The two last terms correspond to a loop touching JS$_I$ or JS$_J$ boundary (Figure \ref{2bl1}).

\FIG{13}{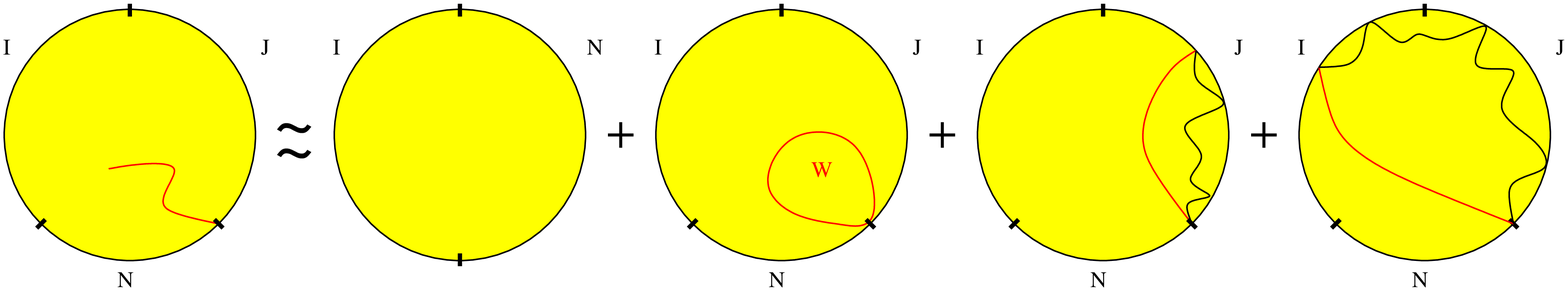}{Combinatoric diagrams describing the splitting of $D_{IJ}\pe$.}{2bl1}

In a similar way, starting with $G=\dfrac{1}{x-X}Y_aH_IH_J$ and summing over $a\in I\cap J$ we derive a second loop equation,
\begin{equation}\label{equ2}
D_{IJ}\pa = k_{IJ} W\ast D_{IJ}\pe+D_{IJ}\pa\ast D_J\pe+D_I\pij\ast D_{IJ}\pe.
\end{equation}
Again, a combinatorial interpretation can be given to each term of the RHS as shown in Figure \ref{2bl2}. We remove the part of the open line starting between Neumann-JS$_J$ boundaries and ending the first time it touches the boundary. This picture corresponds to three terms, either the open line does not touch any JS boundaries, or it first touches JS$_J$ or JS$_I$.
\FIG{14}{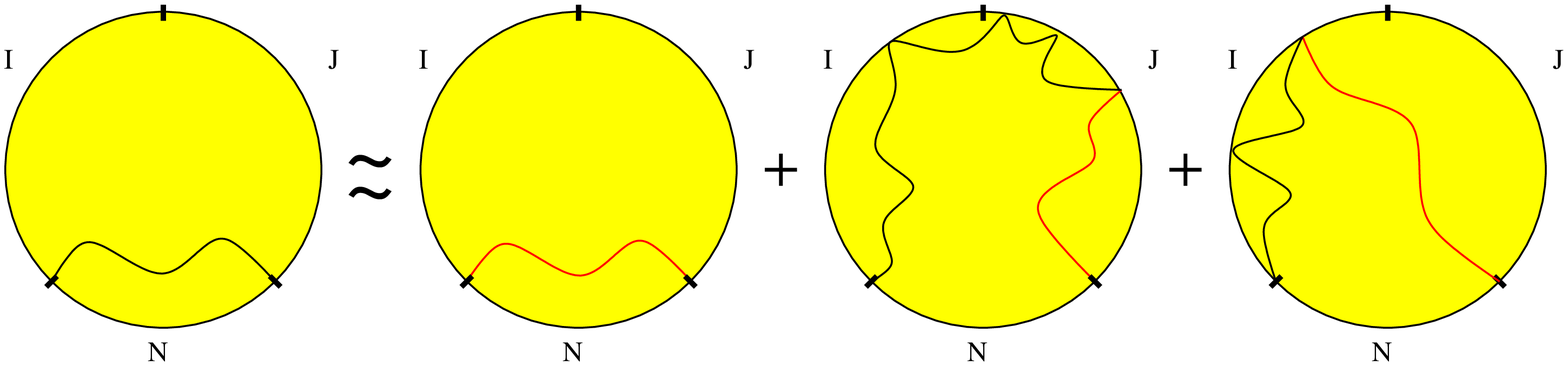}{Combinatoric diagrams describing the splitting of $D_{IJ}\pa$.}{2bl2}

Note that similar equations can be obtained reversing the roles played by $I$ and $J$ boundaries. Using the reflection symmetry of 3-boundaries correlators (\ref{reflection})
we can finally write the following set of coupled loop equations,
\begin{align}
 \begin{split}\label{lenc}
&y_JD_{IJ}\pe=D_{IJ}\pe\ast\left(k_J W +D_J\pa\right)+D_I\pe\ast D_{IJ}\pa+D_I\pe,\\
&D_{IJ}\pa = \left(k_{IJ} W+D_J\pij\right)\ast D_{IJ}\pe+D_{IJ}\pa\ast D_I\pe.
 \end{split}
\end{align}

In order to subtract the non-critical part of 2-boundaries correlators we introduce the quantities \cite{BH:2009},
\begin{align}
\begin{split}\label{redef}
&d_I\pe(x,y_I)=D_I\pe(x,y_I)-1,\\
&k_Id_I\pa(x,y_I)=D_I\pa(x,y_I)+k_IW(x)-y_I,\\
&k_{IJ}d_I\pij(x,y_I)=D_I\pij(x,y_I)+k_{IJ}W(x)-y_I.
\end{split}
\end{align}
The non-critical part of $d_{IJ}\pe$ and $d_{IJ}\pa$ cannot be determined explicitly in general but we do not need their exact expression, let us just denote for simplicity
\begin{equation}
d_{IJ}\pe(x,y_I,y_J)=D_{IJ}\pe(x,y_I,y_J),\quad d_{IJ}\pa(x,y_I,y_J)=D_{IJ}\pa(x,y_I,y_J)+1
\end{equation}
and write \ref{lenc} as
\begin{align}
\begin{split}
&0=d_{IJ}\pe\ast k_Jd_J\pa+d_I\pe\ast d_{IJ}\pa,\\
&0=k_{IJ}d_J\pij\ast d_{IJ}\pe+d_{IJ}\pa\ast d_I\pe.
\end{split}
\end{align}

The quantity $d_I\pij$ that appeared in \ref{equ2} is very similar to $d_I\pa$, they both obeys the equation
\begin{equation}
d_I\pij\ast d_I\pe+W(x)=0.
\end{equation}
This equation implies that $d_I\pij$ and $d_I\pa$ have the same discontinuity along their branch cut, and we can replace $d_I\pij$ by $d_I\pa$ in the previous equation since it appears on the left side of a star product (see Appendix A.2 for more details),\footnote{Two more equations can be obtained simply by exchanging $I$ and $J$.}
\begin{align}
\begin{split}\label{lec}
&d_I\pe\ast d_{IJ}\pa + k_J d_{IJ}\pe\ast d_J\pa =0,\\
&d_{IJ}\pa\ast d_I\pe + k_{IJ}d_J\pa\ast d_{IJ}\pe=0.
\end{split}
\end{align}
This set of two coupled equations is one of the main results of this paper. In the continuum limit, it allows us to determine the gravitational scaling of $d_{IJ}\pe$ hence the dimension of the operator inserted between JS$_I$ and JS$_J$. By taking the sum of these two identities at respectively $x$ and $-x$, we can eliminate one of the star products using the property \ref{comast}. Unfortunately it is not possible to eliminate the second star product in the general case. This can only be done in some specific cases detailed in appendix A.3. These special cases reveal helpfull to remove undeterminations left in the general case.

\subsection{Obtaining the gravitational dimension in the continuum limit}
We take the continuum limit as in \ref{contlim}, \ref{contlim3pts}. Furthermore, we eliminate all but one scaling parameter by imposing critical bulk and JS boundaries, $\m=0=\z_I=\z_J$. Then, every correlator writes as a power of the remaining cosmological constant $\xi$, $d_{IJ}(\xi)=d_{IJ}\xi^{\a_{IJ}}$ where $d_{IJ}$ is an unimportant constant. In this limit, the critical part of the star product between two correlators $d_0(\xi)$ and $d_1(\xi)$ becomes rather trivial and is given by \ref{solstar}. Applied to the loop equations \ref{lec} it gives
\begin{align}
\begin{split}
&d_I\pe d_{IJ}\pa\sin{\pi\a_I\pe}+d_{IJ}\pe d_J\pa k_J\sin{\pi\a_{IJ}\pe}=0,\\
&d_I\pe d_{IJ}\pa\sin{\pi\a_{IJ}\pa}+d_{IJ}\pe d_J\pa k_{IJ}\sin{\pi\a_J\pa}=0.
\end{split}
\end{align}
where $d_I\pe$, $d_J\pa$, $d_{IJ}\pe$ and $d_{IJ}\pa$ are some constants. We recover the relation \ref{relscalings} by imposing both terms to have the same scaling in $\xi$. The ratio of the previous relations gives
\begin{equation}
\dfrac{d_{IJ}\pe d_J\pa}{d_{IJ}\pa d_I\pe}=-\dfrac{\sin{\pi\a_I\pe}}{k_J\sin{\pi\a_{IJ}\pe}}=-\dfrac{\sin{\pi\a_{IJ}\pa}}{k_{IJ}\sin{\pi\a_J\pa}}
\end{equation}
and finally
\begin{equation}\label{kij}
k_{IJ}=\dfrac{\sin{\pi\a_{IJ}\pe}\sin{\pi\a_{IJ}\pa}}{\sin{\pi\a_I\pe}\sin{\pi\a_J\pe}}.
\end{equation}

It is now convenient to introduce the parameterization \ref{paramDJS} of \cite{Dubail:2008}. This parameterization $k_{IJ}(r_{IJ})$ is not well defined because of the periodicity $r_{IJ}\to r_{IJ}+2/\th$, that's why we impose $r_{IJ}\in[1,1+2/\th]$. Relations \ref{kij} and \ref{relscalings} translates into
\begin{equation}
\a_{IJ}\pe=\left(1+r_I+r_J\pm r_{IJ}\right)\dfrac{\th}{2}+j,\quad j\in\mathbb{Z}.
\end{equation}
A look at the results for $k_{IJ}=0$ and $k_{IJ}=k_J$ detailed in appendix A.3 leads to select the plus sign and $j=-1$. This choice provides a dimension $\d_{r_{IJ},r_{IJ}}$ to the boundary operator inserted between the two JS boundaries. This fits in the table of boundary operators dimensions found in \cite{Dubail:2008} on the regular lattice.

\subsection{Loop equations and Liouville theory}
The loop equations \ref{lec} are not sufficient to recover the exact solution for the Liouville boundary 3-points function of \cite{Ponsot:2001ng}. It is nonetheless possible to transform them into the shift relation \ref{prop3pt}. One should also keep in mind the presence of leg factors between matrix model correlators in the continuum limit and boundary Liouville two and three points function. Consequently it makes sense to compare matrix model loop equations and boundary Liouville shift relations only up to an undetermined factor independent of the boundary parameter $\t$. Besides its own interest, the mapping of the matrix model loop equations on Liouville shift relations allows to cross-check the  calculation of the gravitational scalings.

Let us start with the discontinuity of equations \ref{lec} over the support of eigenvalues,
\begin{align}
\begin{split}
&d_{IJ}\pa(-\xi)\Disc d_I\pe+k_Jd_J\pa(-\xi)\Disc d_{IJ}\pe=0,\\
&d_I\pe(-\xi)\Disc d_{IJ}\pa+k_{IJ}d_{IJ}\pe(-\xi)\Disc d_J\pa=0.
\end{split}
\end{align}
This description in terms of discontinuities is totally equivalent to the star product relations, provided we know the behavior of correlators at infinity. In the parameterization \ref{param} these relations become shift equations involving the boundary parameter $\t\in\mathbb{R}$ of the Neumann boundary,
\begin{align}
\begin{split}
&\dfrac{\sin{\pi\p_\t}}{C_I\pe\sinh{b^2\t}}d(P_I\pe,P_{IJ},P_J\pe\mid\s_I,\s_J,\t)\\
=&-\dfrac{d(P_I\pe-b/2\mid\t,\s_I)}{k_Jd(P_J\pa\mid\t,\s_J)}d(P_I\pa,P_{IJ},P_J\pa\mid\s_I,\s_J,\t),\\
&\dfrac{\sin{\pi\p_\t}}{C_J\pa\sinh{b^2\t}}d(P_I\pa,P_{IJ},P_J\pa\mid\s_I,\s_J,\t)\\
=-&\dfrac{d(P_J\pa-b/2,\mid\t,\s_J)}{k_{IJ}d(P_I\pe,\mid\t,\s_I)}d(P_I\pe,P_{IJ},P_J\pe\mid\s_I,\s_J,\t).
\end{split}
\end{align}
Here $C_I\pe$ and $C_J\pa$ are unimportant constants independent of $\t$. To derive this set of equations, we exploited the shift property of boundary Liouville 2-points functions \ref{shift2pt}. If we insert the values of the momenta given by \ref{momenta2pt} we can write
\begin{align}
\begin{split}
&\dfrac{\sin{\pi\p_\t}}{C_I\pe\sinh{b^2\t}}d(P_I\pe,P_{IJ},P_J\pe\mid\s_I,\s_J,\t)\\
=&-\dfrac{d(P_I\pe-b/2\mid\t,\s_I)}{k_Jd(b/2-P_J\pe\mid\t,\s_J)}d(b/2-P_I\pe,P_{IJ},b/2-P_J\pe\mid\s_I,\s_J,\t)
\end{split}
\end{align}
And an equivalent equation under the trivial change of variables $P_{I,J}\pe\to b/2-P_{I,J}\pe$. Finally, the reflection property of boundary operators \ref{reflectBO} bring us the conclusion that the two previous equations coincide with the relation \ref{prop3pt}.

\subsection{A few remarks}
Let us conclude the study of the closed loops sector with a few remarks. First, based on an observation made by Kostov in \cite{Kostov:2007jj} for the disc with Neumann-JS boundaries, we extend the relations \ref{dl} to the case $L=0$. This defines a new quantity $D_{IJ}\paa$ satisfying
\begin{equation}\label{defd0}
D_{IJ}\pa=k_{IJ} W\ast D_{IJ}\paa.
\end{equation}
The compatibility of this definition with \ref{lenc} results in the relation
\begin{equation}\label{reld0}
D_{IJ}\paa(x)=\dfrac{D_{IJ}\pe(x)}{d_I\pe(x)d_J\pe(x)},
\end{equation}
handing only when the image $-x$ belongs to the support of the eigenvalue density. But two functions taking the same values on this symmetric support leads to the same results when appearing on the right side of the star product, so that we have the freedom to choose $D_{IJ}\paa$ such that the relation \ref{reld0} extends to the whole complex plane. This relation becomes significant when written as 
\begin{equation}
D_{IJ}\paa=\dfrac{D_{IJ}\pe}{(1-D_I\pe)(1-D_J\pe)}=\sum_{m,n=0}^\infty{\left(D_I\pe\right)^n\left(D_J\pe\right)^mD_{IJ}\pe}.
\end{equation}
The sum over $(m,n)$ counts the number of time JS$_I$ and JS$_J$ can touch themselves, boundaries touching eachother being forbidden.

A second remark of importance can be made when we compare to the results derived on the regular lattice \cite{Dubail:2008}. Indeed, the spectrum of boundary operators that was found is actually $\d_{r_{IJ}+2j,r_{IJ}}$ with $j\in\mathbb{Z}$ where the weight of loops touching both JS boundaries $k_{IJ}(r_{IJ})$ is parameterized as in \ref{paramDJS} but with no restriction on the range of $r_{IJ}$. This kind of parameterization is invariant under the periodicity transformation
\begin{equation}
r_{IJ}\to r_{IJ}+\dfrac{2}{\th},\quad \d_{r_{IJ}+2j,r_{IJ}}\to\d_{r_{IJ}+2(j+1),r_{IJ}}.
\end{equation}
For concreteness we decided to restrict $r_{IJ}$ to the interval $[1,1+2/\th]$ and we recovered the dimension $\d_{r_{IJ},r_{IJ}}$ ($j=0$) for the JS-JS boundary operator. It is possible to obtain more general boundary operators if we consider the Neumann-JS$_I$-Neumann-JS$_J$ matrix correlator,
\begin{equation}
D_{NINJ}(x,y_I,x',y_J)=\dfrac{1}{\b}\la\tr{\dfrac{1}{x-X}H_I\dfrac{1}{x'-X}H_J}\ra.
\end{equation}
This correlator obeys the same loop equation as $D_{IJ}\pe$ and its critical part have the gravitational dimension, $d_{NINJ}\sim\m^{\g_{IJ}\pe-\frac{1}{2b^2}}$. This dimension can be easily read of the loop equation when $k_{IJ}=0$, the generalization to arbitrary value of $k_{IJ}$ follow the same steps as the three boundary case. When the boundary cosmological constant $x'$ diverges, the length of the corresponding Neumann boundary tends toward zero, JS$_I$-Neumann and Neumann-JS$_J$ boundary operators fuse to form JS$_I$-JS$_J$ boundary operators. This translates into the expansion
\begin{equation}
D_{NINJ}(x,y_I,x',y_J)=\sum_{j=0}^\infty{x^{'-j-1}\ \dfrac{1}{\b}\la\tr{\dfrac{1}{x-X}H_IX^jH_J}\ra}.
\end{equation}
In the continuum limit the Neumann boundary of length $j$ introduced by the matrix product $X^j$ vanish but the presence of this remnant of the random lattice modifies the JS-JS boundary operator. Hence, the matrix model correlator
\begin{equation}
D_{NINJ}^{(j)}(x,y_I,y_J)=\dfrac{1}{\b}\la\tr{\dfrac{1}{x-X}H_IX^jH_J}\ra
\end{equation}
describes in the continuum limit the Neumann-JS$_I$-JS$_J$ disc partition function with a boundary operator of dimension $\d_{r_{IJ}+2j,r_{IJ}}$ inserted between the two JS boundaries.\footnote{It is possible to prove by recursion that such correlators have a gravitational dimension $\g_{NINJ}^{(j)}=\g_{IJ}\pe+\frac{j}{2b^2}$. This is easy to do when $k_{IJ}=0$, we generalize then to arbitrary $k_{IJ}$ arguing that loop equations are the same than in the three boundary case.} Operators of dimension $\d_{r_{IJ}-2j,r_{IJ}}$ where $j$ is a positive integer can be constructed with negative power for the $X$ matrix. However they are not involved in the fusion of Neumann-JS boundary operators. Furthermore, these peculiar operators are wrongly dressed by the Liouville field.

\section{Open lines sector}
We discuss in this section correlators with open lines inserted between JS$_I$ and JS$_J$ boundaries (Figure \ref{2b3}). Amongst these open lines, $L_1$ (resp. $L_2$) will end at the transition Neumann-JS$_I$ (resp. JS$_J$). Such correlators should be compared to the annulus with $L_1+L_2$ non-contractible lines surrounding the inner ring in \cite{Dubail:2008}. On the annulus the first and the last lines can be authorized (or forbidden) to touch the JS boundary, leading to blobbed and unblobbed sectors. Similarly on the disc, the open line next to the JS boundary will be able (or not) to touch this boundary.
\FIG{4}{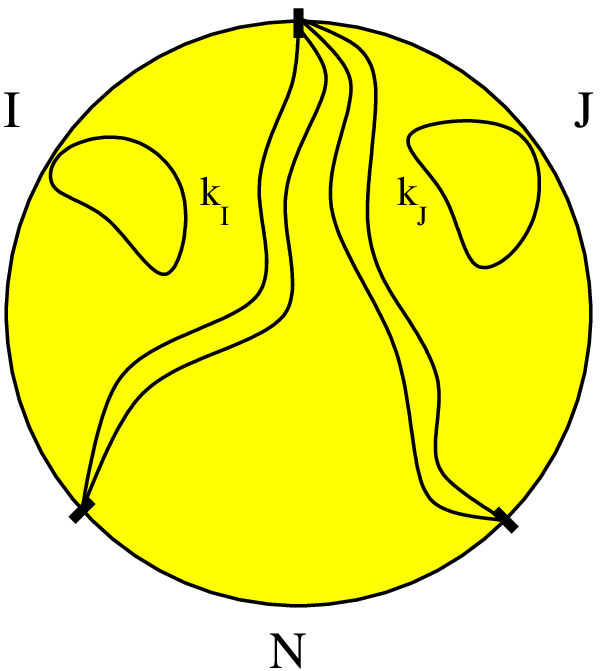}{The disc correlator $D_{IJ}^{(2\ 2\mid E_1\ E_2)}$.}{2b3}

When we remove through the loop equation an open line which is alone, different equations arise due to the possibility of this line to touch the JS boundaries or not. This results in four cases of study corresponding to $(L_1,L_2)\in\{(0,1),(0,L>1),(L>0,1),(L_1>1,L_2>1)\}$.

The presence of open lines inserted between JS$_I$ and JS$_J$ boundaries forbids loops touching both boundaries so that the resulting disc partition function should be independent of the weight $k_{IJ}$. To be more precise, these lines divide the disc into two parts containing respectively JS$_I$ and JS$_J$ boundaries and our results should not rely upon the intersection $I\cap J$. The non-trivial dependence is only on the intersection of the sets describing indices involved in the open lines and the sets characterizing JS boundaries. This gives rise to discussions on blobbed and unblobbed cases with respect to one or two boundaries.

\subsection{$(L>0,1)$ open lines sector}
Let us start with the most rewarding case of $L$ open lines on the JS$_I$ side and only one on the other side. We associate to the JS$_I$ open lines the integer set $E_1$ and $E_2$ to the JS$_J$ open line, considering the correlator $D_{IJ}\le2$. Removing the single JS$_J$ open line, we get
\begin{equation}\label{leL1}
D_{IJ}\le2=k_{E_2 }D_I^{(L\ E_1)}\ast D_J\pe+D_{IJ}^{(L\ 1\mid E_1\ E_2\cap J)}\ast D_J\pe.
\end{equation}
This equation requires two different treatments according to the blob of the open line we remove.

\subsubsection{Unblobbed sector with respect to JS$_J$ boundary}
The simplest case is the unblobbed sector $E_2\cap J=\emptyset$ where the second term of the RHS in the loop equation \ref{leL1} describing the open line touching the JS$_J$ boundary vanishes. Injecting the critical part of $D_I\pe$ yields to
\begin{equation}\label{leL1u}
D_{IJ}\le2=k_{E_2 }D_I^{(L\ E_1)}\ast d_J\pe+k_{E_2}D_I^{(L\ E_1)}(x).
\end{equation}
The second term $D_I^{(L\ E_1)}$ of the RHS cancels with the non-critical part of $D_{IJ}\le2$ as it describes the vanishing of the JS$_J$ boundary. We can read directly the gravitational scaling $\g_{IJ}\le2=\g_I^{(L\ E_1)}+\g_J\pe$. Plugging the values $P_I$ and $P_J$ of the momenta for the Neumann-JS boundary operators into the relation \ref{KPZ3pts} we find out the scaling dimension 
\begin{equation}
\d_{r_J\pm r_I+1,r_J\pm r_I-L}
\end{equation}
for the JS$_I$-JS$_J$ boundary operator with $L+1$ open lines, the plus sign being affected to the JS$_I$ unblobbed open lines case. 

\subsubsection{Blobbed sector with respect to JS$_J$ boundary}
When the removed open line is allowed to touch the JS$_J$ boundary the discussion is more subtle. Indeed, $E_2\subset J$ and the loop equation \ref{leL1} reads,
\begin{equation}\label{leL1b}
D_{IJ}\le2=k_{E_2 }D_I^{(L\ E_1)}\ast D_J\pe+D_{IJ}\le2\ast D_J\pe.
\end{equation}
This equation looks quite similar to the one derived in \cite{BH:2009} for $D_I\pa$,
\begin{equation}
D_I\pa=k_I W\ast D_I\pe+D_I\pa\ast D_I\pe.
\end{equation}
This similarity is not very surprising in the sense that the JS$_I$ boundary is completely decoupled from the JS$_J$ boundary by the $L$ open lines. This JS$_I$ boundary plays no role in the loop equation and the removed open line sees only an effective Neumann boundary condition corresponding to the rightmost JS$_I$ open line. Hence we mimic the solution of the one JS boundary case and take a similar a non-critical part for $D_I\pa$ and $D_{IJ}\le2$,
\begin{align}
\begin{split}
&d_I\pe(x)=D_I\pe(x)-1,\\
&k_Id_I\pa(x)=D_I\pa(x)+k_IW(x)-y_I,\\
&k_{E_2 }d_{IJ}^{(L\ 1\mid E_1\ E_2)}(x)=D_{IJ}\le2(x)+k_{E_2 }\left(D_I^{(L\ E_1)}(x)-c\right).
\end{split}
\end{align}
The exact value of the constant $c$ (independent of $x$) is not important here and will be kept undetermined, the crucial point is to substract the term corresponding to the vanishing of the JS$_J$ boundary. The critical loop equation reads
\begin{equation}
D_I^{(L\ E_1)}+d_{IJ}^{(L\ E_2)}\ast d_J\pe=c.
\end{equation}
and the constant $c$ must cancel with the non-critical part of $D_I^{(L\ E_1)}$. The relation between gravitational scalings is simply $\g_{IJ}^{(L\ 1\mid E_1\ E_2)}=\g_I^{(L\ E_1)}-\g_J\pe$, when plugged in \ref{KPZ3pts} we end up with the scaling dimension
\begin{equation}
\d_{-r_J\pm r_I+1,-r_J\pm r_I-L}
\end{equation}
for the JS$_I$-JS$_J$ boundary operator, the plus sign corresponding to unblobbed open lines with respect to the JS$_I$ boundary.

\vskip 0.7cm
Following the steps of section 3.3 both equations \ref{leL1u} and \ref{leL1b} can be transformed into the shift relation \ref{BHKM2} for the Liouville 3-points function. This confirms our identification of the non-critical part for matrix model correlators.

\subsection{$(0,L>1)$ open lines sector}
This case is very similar to the previous one but we briefly mention how it is solved for completeness. The loop equation reads
\begin{align}\label{le0l}
\begin{split}
D_{I\bar{J}}^{(L\ E)}\ =\ &(k_E-L)\ D_I\pe\ast D_J^{(L-1\ E)}\\
&+\sum_{a_i\in E}{\d_{a_1\in I}D_I\pe\ast \la\tr{\dfrac{1}{x-X}Y_{a_1}H_IY_{a_1}\cdots Y_{a_L}H_JY_{a_L}\cdots Y_{a_2}}\ra},
\end{split}
\end{align}
the second term of the RHS describes the removal of an open line touching the JS$_I$ boundary. It disappears when we consider an unblobbed open line with respect to this boundary, $E\cap I=\emptyset$. We then deduce $\g_{I\bar J}^{(L\ E)}=\g_I\pe+\g_J^{(L\ E)}$ providing the scaling
\begin{equation}
\d_{1+r_I\pm r_J,1+r_I\pm r_J-L}
\end{equation}
for the JS$_I$-JS$_J$ boundary operator, the plus sign being assigned to unblobbed open lines with respect to JS$_J$ boundary.

When the removed open line is unblobbed with respect to the JS$_I$ boundary, $E\subset I$ and the first term of the RHS of \ref{le0l} cancel with the non critical part of the second term as it describes the vanishing of the JS$_I$ boundary. This term is actually $D_I\pa\ast D_J^{(L-2\ E)}$, so that we determine the gravitational scaling $\g_{I\bar J}^{(L\ E)}=\g_I\pe+\g_I\pa+\g_J^{(L-2\ E)}$. It yields to the scaling dimension
\begin{equation}
\d_{1-r_I\pm r_J,1-r_I\pm r_J-L}
\end{equation}
for the operator inserted between the JS boundaries, and again the sign plus corresponds to unblobbed open lines with respect to the JS$_J$ boundary.

\subsection{$(0,1)$ open line sector}
Here we have to distinguish three subcases, the open line being allowed to touch zero, one or two boundaries. We denote by $E\subset[1,n]$ the integer set containing the spin components involved in the open line. The general loop equations can be derived from \ref{le} and, with the shortcut notations of \ref{maindef}, we can write the following set of loop equations,
\begin{align}
\begin{split}\label{les1l}
&D_{I\bar{J}}^{(E)}=k_ED_I\pe\ast D_J\pe+D_I\pe\ast D_{\bar{I}J}^{(E\cap I)}+D_{I\bar{J}}^{(E\cap J)}\ast D_J\pe,\\
&D_{\bar{I}J}^{(E)}=k_ED_J\pe\ast D_I\pe+D_J\pe\ast D_{I\bar{J}}^{(E\cap J)}+D_{\bar{I}J}^{(E\cap I)}\ast D_I\pe.
\end{split}
\end{align}
The second equation is obtained just by reversing the roles played by $I$ and $J$. Summing both equations at points $x$ and $-x$ and using \ref{comast} to get rid of the star products, we end up with
\begin{align}\label{le1l}
\begin{split}
&D_{I\bar{J}}^{(E)}(x)+D_{\bar{I}J}^{(E)}(-x)\\
\ =\ &k_E\ D_I\pe(x)D_J\pe(-x)+D_I\pe(x)D_{\bar{I}J}^{(E\cap I)}(-x)+D_{I\bar{J}}^{(E\cap J)}(x)D_J\pe(-x).
\end{split}
\end{align}
Let us now specialize to the different subcases.

\subsubsection{Unblobbed open line with respect to both boundaries ($E\cap I=\emptyset=E\cap J$)}
When the open line does not touch any of the JS boundaries the star product equations \ref{les1l} simplify into,
\begin{align}
\begin{split}
&D_{I\bar{J}}^{(E)}(x)=k_E\left(d_I\pe\ast d_J\pe\right)(x)+k_ED_I\pe(x),\\
&D_{\bar{I}J}^{(E)}(x)=k_E\left(d_J\pe\ast d_I\pe\right)(x)+k_ED_J\pe(x).
\end{split}
\end{align}
These equations looks very similar to those of section 4.1.1 therefore we take the same kind of non-critical part for the 3-boundaries correlators,
\begin{align}
\begin{split}
&k_Ed_{I\bar{J}}^{(E)}(x)=D_{I\bar{J}}^{(E)}(x)-k_E\left(D_I\pe(x)-c_{I\bar{J}}\right),\\
&k_Ed_{\bar{I}J}^{(E)}(x)=D_{\bar{I}J}^{(E)}(x)-k_E\left(D_J\pe(x)-c_{\bar{I}J}\right).
\end{split}
\end{align}
The loop equation \ref{le1l} implies $c_{I\bar J}+c_{\bar I J}=1$, and by symmetry it is tempting to set $c_{I\bar J}=c_{\bar IJ}=1/2$. We finally get
\begin{equation}\label{le01uu}
d_{I\bar{J}}^{(E)}(x)+d_{\bar{I}J}^{(E)}(-x)=d_I\pe(x)d_J\pe(-x)
\end{equation}
and the relation $\g_{\bar{I}J}^{(E)}=\g_{I\bar{J}}^{(E)}=\g_I\pe+\g_J\pe$ amongst the gravitational scalings. The KPZ relation \ref{KPZ3pts} yield to a scaling 
\begin{equation}
\d_{1+r_I+r_J,r_I+r_J}
\end{equation}
for the operator inserted between the two JS boundaries.

\subsubsection{Blobbed open line on JS$_J$, unblobbed on JS$_I$ ($E\subset J$, $E\cap I=\emptyset$)}
Again, it is more rewarding to consider the loop equation in the start product form \ref{les1l},
\begin{align}
\begin{split}
&D_{I\bar J}\ee(x)=k_E\left(D_I\pe\ast D_J\pe\right)(x)+\left(D_{I\bar J}\ee\ast D_J\pe\right)(x),\\
&D_{\bar I J}\ee(x)=k_E\left(D_J\pe\ast D_I\pe\right)(x)+\left(D_J\pe\ast D_{I\bar J}\ee\right)(x).
\end{split}
\end{align}
The first equation is very similar to the loop equation obtained for $D_I\pa$, so that we define a similar non-critical part that corresponds to the vanishing of the JS$_J$ boundary,
\begin{equation}
k_Ed_{I\bar J}\ee(x)=D_{I\bar J}\ee(x)+k_E\left(D_I\pe(x)-c_{I\bar J}\right).
\end{equation}
This allows us to rewrite the loop equations as
\begin{align}
\begin{split}
&0=\left(d_{I\bar J}\ee\ast d_J\pe\right)(x)+D_I\pe(x),\\
&D_{\bar I J}\ee(x)=k_E\left(d_J\pe\ast d_{I\bar J}\ee\right)(x)+k_Ec_{I\bar J}D_J\pe(x).
\end{split}
\end{align}
The term proportional to $D_J\pe$ should cancel with the non-critical part of $D_{\bar I J}$, so we define
\begin{equation}
k_Ed_{\bar I J}\ee(x)=D_{\bar IJ}\ee(x)-k_E\left(c_{I\bar J}D_J\pe(x)+c_{\bar IJ}\right).
\end{equation}
Such a critical part leads to the relation
\begin{equation}\label{le01bu}
d_{\bar IJ}\ee(-x)=d_{I\bar J}\ee(x)d_J\pe(-x)+d_I\pe(x)-c_{I\bar J}-c_{\bar IJ}+1.
\end{equation}
We get $c_{I\bar J}+c_{\bar IJ}=1$, and by analogy with respect to the first section, it is tempting to set $c_{I\bar J}=c_{\bar IJ}=1/2$. The gravitational scalings are related by $\g_{\bar IJ}\ee=\g_I\pe=\g_{I\bar J}\ee+\g_J\pe$ and both leads to a scaling dimension
\begin{equation}
\d_{1+r_I-r_J,r_I-r_J}
\end{equation}
for the operator inserted between the JS boundaries.

\subsubsection{Blobbed open line on both JS boundaries ($E\subset I\cap J$)}
We start from the loop equation \ref{le1l},
\begin{equation}
D_{I\bar{J}}^{(E)}(x)+D_{\bar{I}J}^{(E)}(-x)=k_ED_I\pe(x)D_J\pe(-x)+D_I\pe(x) D_{\bar{I}J}^{(E)}(-x)+D_{I\bar{J}}^{(E)}(x)D_J\pe(-x).
\end{equation}
By analogy with the previous case, we define our critical quantities as
\begin{align}
\begin{split}
&k_Ed_{I\bar J}^{(E)}(x)=D_{I\bar J}^{(E)}(x)+k_E\left(\frac{1}{2}D_I\pe(x)+\frac{1}{2}\right),\\
&k_Ed_{\bar IJ}^{(E)}(x)=D_{\bar IJ}^{(E)}(x)+k_E\left(\frac{1}{2}D_J\pe(x)+\frac{1}{2}\right).
\end{split}
\end{align}
thus obtaining the critical loop equation
\begin{equation}\label{le01bb}
0=1+d_I\pe(x)d_{\bar IJ}\ee(-x)+d_J\pe(-x)d_{I\bar J}\ee(x)
\end{equation}
leading to gravitational scalings $\g_{\bar IJ}=-\g_I\pe$, $\g_{I\bar J}=-\g_J\pe$, the scaling dimension of the operator inserted between the two JS boundaries reads
\begin{equation}
\d_{1-r_I-r_J,-r_I-r_J}.
\end{equation}
This case is the only one in the open line sector for which the momentum of the Liouville dressing factor is negative. As explained in the first section, in such a case we have to use a wrong dressing of the bare boundary operator.

\vskip 0.7cm
To conclude the study of the $(0,L>1)$ sector, let us mention that the three critical loop equations \ref{le01uu},\ref{le01bu},\ref{le01bb} lead to the relation \ref{BHKM1} after a few algebraic manipulation. This strong result ensures the correctness of our ansatz for the non-critical parts of the 3-boundaries matrix correlators.

\subsection{$(L_1>1,L_2>1)$ open lines sector}
This case is rather trivial because the line we remove is provided to touch JS boundaries by the other open lines. In this way we do not need to specialize to any blobbed/unblobbed sector and the loop equation simply express the splitting of the disc into two parts, \begin{equation}\label{lefin}
D_{IJ}^{(L_1\ L_2\mid E_1\ E_2)}=k_{E_2}\ D_I^{(L_1\ E_1)}\ast D_J^{(L_2-1\ E_2)}.
\end{equation}
The gravitational scaling of the LHS is just the sum of the scalings for both RHS correlators. The momenta of Neumann-JS boundary operators are given by
\begin{equation}
P_I=\e_Ir_I\left(\dfrac{1}{2b}-\dfrac{b}{2}\right)+L_1\dfrac{b}{2},\quad P_J=\e_Jr_J\left(\dfrac{1}{2b}-\dfrac{b}{2}\right)+L_2\dfrac{b}{2}
\end{equation}
where $\e=\pm 1$ is a sign corresponding to the blob of the open lines, $\e=+1$ in the unblobbed case. We then derive the scaling dimension 
\begin{equation}
\d_{\e_Ir_I+\e_Jr_J+1,\e_Ir_I+\e_Jr_J+1-(L_1+L_2)}
\end{equation}
for the JS$_I$-JS$_J$ boundary operator. From our results of the previous sections, we can extend this formula to any values of $L_1$ and $L_2$. These results are in agreement with the analysis made in \cite{Dubail:2008} on the regular lattice, provided we identify the number of non contractible loops of the annulus to the total number of open lines $L_1+L_2$. Note that all these results are, as expected, independent of the intersection $I\cap J$. It is also satisfactory that the scaling of the operator depends only on the total number of open lines $L_1+L_2$, being insensitive to where these open lines end.

As in the closed loops sector, it is possible to construct more general boundary operators with scaling dimension
\begin{equation}
\d_{\e_Ir_I+\e_Jr_J+1+2j,\e_Ir_I+\e_Jr_J+1-(L_1+L_2)},\quad j\in\mathbb{Z}_+.
\end{equation}
This is done in a similar way, introducing a product of $j$ $X$ matrices between the two JS boundaries. We then obtain the same kind of loop equations, they correspond to the shift relation \ref{BHKM2} in the continuum limit, where the momentum of the JS-JS boundary operator is 
\begin{equation}
P=P_I+P_J+\dfrac{1}{2b}(2j+1)-\dfrac{b}{2},\quad j\in\mathbb{Z}_+.
\end{equation}

\section{Concluding remarks}
% We can now address the most general case of the disc correlator with $L_1$ open lines inserted from Neumann-JS$_I$ to JS$_I$-JS$_J$, $L_2$ open lines from Neumann-JS$_J$ to JS$_I$-JS$_J$, and $L_3$ open lines from Neumann-JS$_I$ to Neumann-JS$_J$,
% \begin{equation}
% D_{IJ}^{(L_1\ L_2\ L_3)}(x)=\dfrac{1}{\b}\la\tr{\dfrac{1}{x-X}Y_{L_3}^{(E_3)}Y_{L_1}^{(E_1)}H_IY_{L_1}^{(E_1)\dagger}Y_{L_2}^{(E_2)}H_JY_{L_2}^{(E_2)\dagger}Y_{L_3}^{(E_3)\dagger}}\ra
% \end{equation}
% with the no-bounce requirement $E_i\cap E_j=\emptyset$. When $L_3\geq 1$ we easily derive the recursive loop equations
% \begin{equation}\label{3ol}
% D_{IJ}^{(L_1\ L_2\ L_3)}(x,y_I,y_J)=(k_{E_3}-L)W\ast D_{IJ}^{(L_1\ L_2\ L_3-1)}(x)
% \end{equation}
% where $k_{E_3}=\Card E_3$. This equation simply means that the operator inserted between both JS boundaries remain unaffected by the presence of the $L_3$ open lines Neumann-JS.

The main results of the present article are formulas \ref{res1} and \ref{res2} for the scaling dimension of JS-JS boundary operators in closed loops and open lines sectors. We provide an independent check of the results obtained in \cite{Dubail:2008} using numerics and Coulomb gas arguments. Our method can be easily generalized to more complicated topologies with non-trivial cycles. This matrix model approach carries many interesting features. For instance the fusion of two Neumann-JS boundary operators can be done explicitly in sending the boundary cosmological constant toward infinity. In the closed loop sector, the fusion rules depends not only on the JS parameters $k_I$ and $k_J$ but also on the symmetry group that is preserved by both boundaries, through the weight $k_{IJ}$ of loops touching both JS boundaries. At fixed $k_{IJ}$, the fusion rules contain an infinite number of terms,
\begin{equation}
\d_{r_I,r_I}\times\d_{r_J,r_J}=\bigoplus_{j=0}^\infty{\d_{r_{IJ}+2j,r_{IJ}}}
\end{equation}
where $r_{IJ}$ is related to $r_I$ and $r_J$ via the parameterization \ref{paramDJS}. In the open line sector, loops touching both boundaries are forbidden and the fusion rules only depends on $k_I$ and $k_J$,
\begin{equation}
\d_{\e_Ir_I,\e_Ir_I-L_1}\times\d_{\e_Jr_J,\e_Jr_J-L_2}=\bigoplus_{j=0}^\infty{\d_{1+\e_Ir_I+\e_Jr_J+2j,1+\e_Ir_I+\e_Jr_J-(L_1+L_2)}}.
\end{equation}
% 
% The $O(n)$ model coupled to gravity can be seen as a string theory with a discrete target space \cite{Kostov:1991cg}. In this framework, the spectrum of conformal boundary conditions (CBC) lists the D-branes of the theory. The dimension of the boundary operators inserted between two CBC gives the string open modes that can propagate between two D-branes. Furthermore, the fusion relations derived above give restrictions on the non-vanishing $S$-matrix entries for scatterings involving three open strings. Pushing forward this analysis would require to understand the dilute phase of the $O(n)$ model. However, we hope to address this issue in a near future, and in particular the $c=1$ limit ($n\to 2$) may show some interesting features.
% Despite the difference of topology all the findings of the flat lattice translate to the random lattice through the KPZ relation. 

As a consistency check, we were able to map the critical loop equations on the boundary ground ring relations obtained in Liouville theory. It would be interesting to develop further this mapping, and in particular to investigate the role of Liouville degenerate boundary operator in the matrix model. Furthermore, a connection between the KPZ relation and Schramm-Loewner evolution (SLE) was found recently in \cite{Duplantier:2009}, \cite{Bauer:2008}. It seems natural to ask for an interpretation of the loop equations in the mathematical framework of SLE.

Finally, for special values of $n$ the continuum limit of the $O(n)$ model is given by a minimal model. Then the conformal matter can be described using Coulomb gas techniques. This description imposes severe restrictions on the Liouville momenta \cite{Furlan:2008}. On the other side, the $O(n)$ model is known to be mapped on a Restricted Solid On Solid (RSOS) model, JS boundary conditions being transpose to alternating height boundary conditions \cite{Jacobsen:2006bn,BH:2009}. It would be very intersting to compare the allowed weight a loop touching two different boundaries with alternating height boundary conditions can take and the restrictions coming from Coulomb gas predictions. Furthermore, one could rederive the fusion rules of minimal models in this context.

\section*{Acknowledgments}

It is a pleasure to thank the hospitality of the NBI where a part of this work has been achieved. The author acknowledges J. Dubail and C. Kristjansen for useful discussion and is debtfull to K. Hosomichi and I. Kostov for their carefull reading of the manuscript.

\appendix
\section{Appendix}
\subsection{Simplification of the shift relations for the boundary Liouville 3-points function}
Using the operator product expansion of boundary ground ring operator, a shift relation for the boundary 3-points function can be derived (see \cite{Ponsot:2001ng,BHKM:2007}):
\begin{align}
\begin{split}\label{BHKM}
&\dfrac{\G(1+2bP_I-\a)}{\G(2bP_I)}\ d(P_I+\frac{b}{2},P,P_J\mid\s_I,\s_J,\t\pm i\pi)\\
-&\dfrac{\G(1-2bP_J)}{\G(\a-2bP_J)}\ d(P_I,P,P_J+\frac{b}{2}\mid\s_I,\s_J,\t)\\
=&\dfrac{\G(1+2bP_J)\G(1+2bP_I-\a)}{\G(1+2bP_I+2bP_J-\a)\G(\a)}\dfrac{d(P_J\mid\s_J,\t\pm i\pi)}{d(P_J-\frac{b}{2}\mid\s_J,\t)}\ d(P_I,P,P_J-\frac{b}{2}\mid\s_I,\s_J,\t)
\end{split}
\end{align}
with $\a=\frac{1}{2}+b(P_I+P_J+P)$.

Taking the difference of shifts with plus and minus sign and using \ref{shift2pt} we recover the relation of \cite{Kostov:2003uh},
\begin{equation}\label{prop3pt}
\dfrac{\sin{\pi\p_\t}}{\sinh{b^2\t}}d(P_I,P,P_J\mid\s_I,\s_J,\t)=\mathcal{C}\ d(P_I-b/2,P,P_J-b/2\mid\s_I,\s_J,\t)
\end{equation}
where $\mathcal{C}$ is some constant depending only on the Liouville momenta.

In \cite{BHKM:2007} the momenta conservation $\a=0$ was taken into account in order to simplify \ref{BHKM}. Simplifications with more general momenta conservations are discussed in \cite{Furlan:2008}. Here we need only to investigate the case $1+2bP_I+2bP_J-\a=0$, i.e.
\begin{equation}
P=\dfrac{1}{2b}+P_I+P_J.
\end{equation}
The zero arising from the gamma function in the RHS of \ref{BHKM} cancels with the pole of the boundary 3-points function,
\begin{equation}
d(P_I,P,P_J-\frac{b}{2}\mid\s_I,\s_J,\t)\sim\text{cst } \dfrac{d(P_J-\frac{b}{2}\mid\s_J,\t)d(P_I\mid\s_I,\t)}{\frac{1}{2b}+P_I+P_J-P}
\end{equation}
where the constant is independent of $\t$. The equation \ref{BHKM} simplifies into
\begin{align}\label{BHKM1}
\begin{split}
&2P_Jd(P_I,P_I+P_J,P_J+\frac{b}{2}\mid\s_I,\s_J,\t)+2P_Id(P_I+\frac{b}{2},P,P_J\mid\s_I,\s_J,\t\pm i\pi)\\
=&\ \mathcal{C}'\ d(P_J\mid\s_J,\t\pm i\pi)d(P_I\mid\s_I,\t).
\end{split}
\end{align}

Taking the difference of both $+i\pi$ and $-i\pi$ shifts of the boundary parameter $\t$, after a convenient change of variable we end up with
\begin{equation}\label{BHKM2}
\dfrac{\sinh{\pi\p_\t}}{\sinh{b^2\t}}\ d(P_I,P_I+P_J+\frac{e_0}{2},P_J\mid\s_I,\s_J,\t)
=\mathcal{C}''d(P_J-\frac{b}{2}\mid\s_J,\t)d(P_I-\frac{b}{2}\mid\s_I,\t)
\end{equation}
where $\mathcal{C}''$ is independent of the boundary parameters and $e_0=\frac{1}{b}-b$. A similar formula can be obtained for the boundary 3-point function $d(P_I,P,P_J,\s_I,\s_J,\t)$ with a momentum
\begin{equation}
P=P_I+P_J+\dfrac{e_0}{2}+\dfrac{j}{b},\quad j\in\mathbb{Z_+}
\end{equation}
between the two JS boundaries.

\subsection{Properties of the star product}
\subsubsection{Main properties}
In this section we investigate some useful properties of the star product $\ast$ previously introduced in \cite{BH:2009} to describe the Laplace transformed convolution. Let us first recall its definition,
\begin{equation}\label{defstar}
\left(A\ast B\right) (x) = \oint_{[a,b]}{\dfrac{dx'}{2i\pi}\dfrac{A(x')-A(x)}{x-x'}B(-x')}
\end{equation}
where the contour circles the support $[a,b]\subset\mathbb{R}^-$ of the eigenvalue density.

We will apply this star product to a restricted set of functions with a single branch cut on the set $[a,b]$, no poles in $\mathbb{C}\setminus [a,b]$ and a constant behavior $A(x)\sim a_0+O(1/x)$ at infinity. This product is bilinear but not symmetric, e.g. the action of any polynome $P(x)$ on the right leads to a vanishing result $P\ast A=0$ whereas the action on the left extract the behavior of $A$ at infinity, for instance:
\begin{equation}
 \left(A\ast 1\right)(x)=A(x)-a_0.
\end{equation}

Deforming the contour of integration on the sphere, the integral \ref{defstar} gives two contributions corresponding to the singularity at infinity and the branch cut of $B(-x)$ located on $\mathbb{R}^+$. This trick allows us to establish the important relation
\begin{equation}\label{comast}
\left(A\ast B\right)(x)+\left(B\ast A\right)(-x)=A(x)B(-x)-a_0b_0.
\end{equation}

The star product has the following discontinuity crossing the branch cut $[a,b]$,
\begin{equation}
\Disc \left(A\ast B\right)(x)=B(-x)\Disc A(x)
\end{equation}
where we denoted
\begin{equation}
\Disc A = A(x+i0)-A(x-i0).
\end{equation}

It is sometimes convenient to replace the contour integral by a usual integral over the branch cut,
\begin{equation}\label{ointdisc}
\left(A\ast B\right)(x)=\int_a^b{\dfrac{dx'}{2i\pi}\dfrac{B(-x')}{x'-x}\text{Disc}_{x'}\ A}.
\end{equation}
We can use this description to prove a few basic properties. If two functions $A_1$ and $A_2$ have the same discrepancy crossing the branch cut then $A_1\ast B=A_2\ast B$. Similarly, if two functions $B_1$ and $B_2$ take the same values on the symmetric support $[-b,-a]\subset\mathbb{R}^+$ then $A\ast B_1=A\ast B_2$.

\subsubsection{The star product in the critical limit}
Let us consider two correlators $D_0(x)$, $D_1(x)$ in the continuum limit,
\begin{equation}
\e^{\a_i}d_i(\xi)=D_i(x)-D_i^*(x)
\end{equation}
where $x=\e\xi$ and $D_i^*$ stands for the non-critical part of $D_i$. In the continuum limit, correlators have a branch cut along the interval $]-\infty,0]$. The star product of $D_0$ and $D_1$ can be written as the sum of a term of order $\e^{\a_0+\a_1}$ corresponding to the star product of $d_0$ and $d_1$ plus some higher order terms involving the non-critical part and the behavior of the correlators at infinity:
\begin{equation}
\e^{\a_0+\a_1}\left(d_0\ast d_1\right)(\xi)=\left(D_0\ast D_1\right)(x)-\left(D_0\ast D_1\right)^*(x).
\end{equation}
The star product of the critical part being given by
\begin{equation}
\left(d_0\ast d_1\right)(\xi)=\oint_{]-\infty,0]}{\dfrac{d\xi'}{2i\pi}\dfrac{d_0(\xi')-d_0(\xi)}{\xi-\xi'}d_1(-\xi')}.
\end{equation}

At the critical point where only the $\xi$ boundary cosmological constant remains, critical parts of the correlators  simply become $d_i(\xi)=d_i\xi^{\a_i}$ and we can easily compute their star product. Indeed, transforming the contour integral using \ref{ointdisc} we get
\begin{equation}
\left(d_0\ast d_1\right)(\xi)=-\dfrac{d_0d_1}{\pi}\sin{\pi\a_0}\ \int_0^\infty{\dfrac{\xi'^{\a_0+\a_1}}{\xi+\xi'}d\xi'}
\end{equation}
since $\Disc \xi^\a=2i\sin{\pi\a}\ (-\xi)^\a$. The last integral depends only on the sum $\a_0+\a_1$ and can be computed using the identity
\begin{equation}
\dfrac{1}{x+a}=\int_0^\infty{dl e^{-lx} e^{-la}}.
\end{equation}
We write
\begin{equation}
\int_0^\infty{\dfrac{\xi'^{\a_0+\a_1}}{\xi+\xi'}d\xi'}=\int_0^\infty{d\xi'\int_0^\infty{dl \ \xi'^{\a_0+\a_1}e^{-l\xi}e^{-l\xi'}}}
\end{equation}
and perform the change of variable $\xi'\to t=\xi' l$. It leads to the appearance of a gamma function,
\begin{equation}
\int_0^\infty{\dfrac{\xi'^{\a_0+\a_1}}{\xi+\xi'}d\xi'}=\G(1+\a_0+\a_1)\int_0^\infty{dl\  l^{-1-\a_0-\a_1} e^{-l\xi}}.
\end{equation}
A second change of variable $l\to l\xi$ finally gives
\begin{equation}
\int_0^\infty{\dfrac{\xi'^{\a_0+\a_1}}{\xi+\xi'}d\xi'}=\G(1+\a_0+\a_1)\G(-\a_0-\a_1)\xi^{\a_0+\a_1}
\end{equation}
Using the identity
\begin{equation}
\G(1-x)\G(x)=\dfrac{\pi}{\sin{\pi x}},
\end{equation}
we end up with
\begin{equation}\label{solstar}
\left(d_0\ast d_1\right)(\xi)=\dfrac{\sin{\pi\a_0}}{\sin{\pi(\a_0+\a_1)}}\ d_0d_1\xi^{\a_0+\a_1}.
\end{equation}
Note that this quantity satisfies the relation
\begin{equation}
\left(d_0\ast d_1\right)(\xi)+\left(d_1\ast d_0\right)(-\xi)=d_0(\xi)d_1(-\xi).
\end{equation}

\subsection{Special cases}
In this appendix we consider several cases for which the loop equation simplify. We are then able to compute their solution in the continuum limit. The values found for the critical exponents are necessary to determine the correct dimension of boundary operators in the general situation.

\subsubsection{$I\cap J=\emptyset$, $k_{IJ}=0$}
When $k_{IJ}=0$, the loop equation of $D_{IJ}\pe$ is self consistent and we do not need to consider $D_{IJ}\pa$ anymore,
\begin{equation}
0=D_{IJ}\pe\ast d_J\pa+D_I\pe.
\end{equation}
In the continuum limit we deduce the gravitational scaling
\begin{equation}
\g_{IJ}\pe=\g_I\pe-\g_J\pa=\g_J\pe-\g_I\pa
\end{equation}
for the operator inserted between both JS boundaries. The equation $k_{IJ}(r_{IJ})=0$ has for solutions
\begin{equation}
r_{IJ}=\pm(1+r_I+r_J)+\dfrac{2j}{\th},\quad j\in\mathbb{Z}.
\end{equation}
If we impose the condition $r_{IJ}\in[1,1+2/\th]$, we get $r_{IJ}=1+r_I+r_J$ and $\g_{IJ}\pe=r_{IJ}\frac{\th}{2b^2}-\frac{1}{2b^2}$, the dimension of the corresponding JS-JS boundary operator is $\d_{r_{IJ},r_{IJ}}$.

\subsubsection{$I=J$, $k_I=k_J=k_{IJ}$}
When $I=J$ correlators simplify into
\begin{align}
\begin{split}
&D_{IJ}\pe(x,y_I,y_J)=\dfrac{D_I\pe(x,y_J)-D_I\pe(x,y_I)}{y_I-y_J}\\
&D_{IJ}\pa(x,y_I,y_J)=\dfrac{D_I\pa(x,y_J)-D_I\pa(x,y_I)}{y_I-y_J}
\end{split}
\end{align}
and in the continuum limit,
\begin{align}
\begin{split}
&d_{IJ}\pe(\xi,\z_I,\z_J)=\dfrac{d_I\pe(\xi,\z_J)-d_I\pe(\xi,\z_I)}{\z_I-\z_J}\\
&d_{IJ}\pa(\xi,\z_I,\z_J)=k_I\dfrac{d_I\pa(\xi,\z_J)-d_I\pa(\xi,\z_I)}{\z_I-\z_J}
\end{split}
\end{align}
where the non-critical parts were found to be $D_{IJ}^{(0\ \perp)*}(x,y_I,y_J)=0$ and $D_{IJ}^{(1\ \parallel)*}(x,y_I,y_J)=-1$. We easily read the scaling dimensions $\g_{IJ}\pe=\g_I\pe-\frac{1}{2}$ and $\g_{IJ}\pa=\g_I\pa-\frac{1}{2}$. The boundary operator introduced between the two JS boundaries must be the identity operator $\d_{1,1}$. If we choose the value $r_{IJ}=1\in[1,1+2/\th]$ for the solution of $k_{IJ}(r_{IJ})=k_{I}=k_J$, the dimension of the JS-JS boundary operators writes $\d_{r_{IJ},r_{IJ}}$.\footnote{There are actually three solutions of $k_{IJ}(r_{IJ})=k_{I}=k_J$ belonging to $[1,1+2/\th]$, namely $r=1$ and $r=\pm1+2/\th$. The choice of $r=1$ is motivated by the study of the general case $k_{IJ}=k_I\neq k_J$.} Note also the relation
\begin{equation}
D_{IJ}\paa(x,y_I,y_J)=\dfrac{D_I\paa(x,y_J)-D_I\paa(x,y_I)}{y_I-y_J}
\end{equation}
compatible with the equation \ref{reld0}.

\subsubsection{$J\subset I$, $k_{IJ}=k_J$}
When $I\cap J=J$ (or $k_{IJ}=k_J$) the star product can be eliminated using the property \ref{comast} applied to the sum of the two equations \ref{lec} at points respectively $x$ and $-x$. We obtain
\begin{equation}\label{spec}
1+d_I\pe(x)\left(1+D_{IJ}\pa(-x)\right)+k_JD_{IJ}\pe(x)d_J\pa(-x)=0.
\end{equation}
In the limit $k_I\to k_J$ we retrieve the previous section case. A study of the loop equation \ref{spec} in this limit leads us to define the same non-critical part for $D_{IJ}\pe$ and $D_{IJ}\pa$ as before. Then, the continuum limit of \ref{spec} reads
\begin{equation}
1+d_I\pe(\xi)d_{IJ}\pa(-\xi)+k_Jd_{IJ}\pe(\xi)d_J\pa(-\xi)=0.
\end{equation}
It follows $\g_{IJ}\pe=-\g_J\pa$ and the JS-JS boundary operator has dimension $\d_{1+r_J-r_I,1+r_J-r_I}$. The equation $k_{IJ}(r_{IJ})=k_J$ as for solutions
\begin{equation}
r_{IJ}=\pm(1+r_J-r_I)+\dfrac{2j}{\th},\quad j\in\mathbb{Z}.
\end{equation}
Restricting to $r_{IJ}\in[1,1+2/\th]$ and taking into account that the set $J$ is included in $I$, i.e. $r_I<r_J$, the only solution is $r_{IJ}=1+r_J-r_I$, in agreement with the formula $\d_{r_{IJ},r_{IJ}}$ for the dimension of the JS-JS boundary operator.


\begin{thebibliography}{99}

\bibitem{Kostov:1991cg}
  I.~K.~Kostov,
  {\it ``Strings with discrete target space,''}
  Nucl.\ Phys.\  B {\bf 376}, 539 (1992)
  {\tt [hep-th/9112059]}.

\bibitem{Kazakov:1986hy}
  V.~A.~Kazakov,
  {\it ``Exact solution of the Ising model
         on a random two-dimensional lattice,''}
  JETP Lett.\  {\bf 44}, 133 (1986)
  [Pisma Zh.\ Eksp.\ Teor.\ Fiz.\  {\bf 44}, 105 (1986)].

\bibitem{Boulatov:1986sb}
  D.~V.~Boulatov and V.~A.~Kazakov,
  {\it ``The Ising model on random planar lattice: the structure of
   	 phase transition and the exact critical exponents,''}
  Phys.\ Lett.\  {\bf 186B}, 379 (1987).

\bibitem{Duplantier:1988wc}
  B.~Duplantier and I.~Kostov,
  {\it Conformal spectra of polymers on a random surface,''}
  Phys.\ Rev.\ Lett.\  {\bf 61}, 1433 (1988);
  {\it Geometrical critical phenomena on a random surface of arbitrary genus,''}
  Nucl.\ Phys.\  B {\bf 340}, 491 (1990).

\bibitem{Kazakov:1988fv}
  V.~K.~Kazakov,
  {\it ``Percolation on a fractal with the statistics
         of planar Feynman graphs: exact solution,''}
  Mod.\ Phys.\ Lett.\  A {\bf 4}, 1691 (1989).

\bibitem{Kostov:1988fy}
  I.~K.~Kostov,
  {\it ``$O(n)$ vector model on a planar random lattice:
         spectrum of anomalous dimensions,''}
  Mod.\ Phys.\ Lett.\  A {\bf 4}, 217 (1989).

\bibitem{Kazakov:1991pt}
  V.~A.~Kazakov and I.~K.~Kostov,
  {\it ``Loop gas model for open strings,''}
  Nucl.\ Phys.\  B {\bf 386}, 520 (1992)
  {\tt [hep-th/9205059]}.

\bibitem{Kostov:2003uh}
  I.~K.~Kostov, B.~Ponsot and D.~Serban,
  {\it ``Boundary Liouville theory and 2D quantum gravity,''}
  Nucl.\ Phys.\  B {\bf 683}, 309 (2004)
  {\tt [hep-th/0307189]}.

\bibitem{Kostov:2002uq}
  I.~K.~Kostov,
  {\it ``Boundary correlators in 2D quantum gravity:
         Liouville versus discrete approach,''}
  Nucl.\ Phys.\  B {\bf 658}, 397 (2003)
  {\tt [hep-th/0212194]}.

\bibitem{Nienhuis:1982fx}
  B.~Nienhuis,
  {\it ``Exact critical point and critical exponents of
         $O(n)$ models in two-dimensions,''}
  Phys.\ Rev.\ Lett.\  {\bf 49}, 1062 (1982).

\bibitem{Nienhuis:1984wm}
  B.~Nienhuis,
  {\it ``Critical behavior of two-dimensional spin models
         and charge asymmetry in the Coulomb gas,''}
  J.\ Statist.\ Phys.\  {\bf 34}, 731 (1984)
  and in
  C.~Domb and J.~L.~Lebowitz,
  {\it ``Phase transitions and critical phenomena. vol. 11,''}
  London, Uk: Academic (1987) 210p.

\bibitem{Kostov:2006ry}
  I.~K.~Kostov,
  {\it ``Thermal flow in the gravitational $O(n)$ model,''}
  {\tt [hep-th/0602075]}.

\bibitem{DiFrancesco:1999}
 P.~Di Francesco, E.~Guitter, C.~Kristjansen
 {\it ``Fully Packed $O(n=1)$ Model on Random Eulerian Triangulations,''}
 Nucl.\ Phys.\ B {\bf 549}, 657 (1999)
{\tt [hep-th/9902082]}.

\bibitem{Jacobsen:2006bn}
  J.~L.~Jacobsen and H.~Saleur,
  {\it ``Conformal boundary loop models,''}
  Nucl.\ Phys.\  B {\bf 788}, 137 (2008)
  {\tt [math-ph/0611078]}.

\bibitem{Nichols:2004fb}
  A.~Nichols, V.~Rittenberg and J.~de Gier,
  {\it ``One-boundary Temperley-Lieb algebras in the XXZ and loop models,''}
  J.\ Stat.\ Mech.\  {\bf 0503}, P003 (2005)
  {\tt [cond-mat/0411512]};
  A.~Nichols,
  {\it ``The Temperley-Lieb algebra and its generalizations
         in the Potts and XXZ models,''}
  J.\ Stat.\ Mech.\  {\bf 0601}, P003 (2006)
  {\tt [hep-th/0509069]};
  A.~Nichols,
  {\it ``Structure of the two-boundary XXZ model with
         non-diagonal boundary terms,''}
  J.\ Stat.\ Mech.\  {\bf 0602}, L004 (2006)
  {\tt [hep-th/0512273]}.

\bibitem{Pearce:2006sz}
  P.~A.~Pearce, J.~Rasmussen and J.~B.~Zuber,
  {\it ``Logarithmic minimal models,''}
  J.\ Stat.\ Mech.\  {\bf 0611}, P017 (2006)
  {\tt [hep-th/0607232]}.

\bibitem{Jacobsen:2008}
J.~Jacobsen and H.~Saleur
{\it ``Combinatorial aspects of boundary loop models,''}
J. Stat. Mech. {\bf 2008}, 01
{\tt [math-ph/0709.0812]}

\bibitem{Kostov:2007jj}
  I.~Kostov,
  {\it ``Boundary loop models and 2D quantum gravity,''}
  J.\ Stat.\ Mech.\  {\bf 0708}, P08023 (2007)
  {\tt [hep-th/0703221]}.

\bibitem{BH:2009}
 J-E.~Bourgine and K.~Hosomichi
 {\it ``Boundary operators in the $O(n)$ and RSOS matrix model,''}
 JHEP {\bf 0111}, 009 (2009)
  {\tt [hep-th/0811.3252]}.

\bibitem{Dubail:2008}
  J.~ Dubail, J.~L.~Jacobsen and H.~Saleur,
  {\it ``Conformal two-boundary loop model on the annulus,''}
  {\tt [arXiv:0812.2746].}

\bibitem{Fateev:2000ik}
  V.~Fateev, A.~B.~Zamolodchikov and A.~B.~Zamolodchikov,
  {\it ``Boundary Liouville field theory. I:
         Boundary state and boundary two-point function,''}
  {\tt [hep-th/0001012]}.

\bibitem{Ponsot:2001ng}
  B.~Ponsot and J.~Teschner,
  {\it ``Boundary Liouville field theory: Boundary three point function,''}
  Nucl.\ Phys.\  B {\bf 622}, 309 (2002)
  {\tt [hep-th/0110244]}.

\bibitem{Hosomichi:2001xc}
  K.~Hosomichi,
  {\it ``Bulk-boundary propagator in Liouville theory on a disc,''}
  JHEP {\bf 0111}, 044 (2001)
  {\tt [hep-th/0108093]}.

\bibitem{Alexandrov:2005}
S.~Alexandrov and E.~Imeroni
{\it ``$c=1$ from $c<1$: Bulk and boundary correlators,''}
Nucl. Phys. B {\bf 731}, 242 (2005)
{\tt [hep-th/0504199]}

\bibitem{Knizhnik:1988ak}
  V.~G.~Knizhnik, A.~M.~Polyakov and A.~B.~Zamolodchikov,
  {\it ``Fractal structure of 2d-quantum gravity,''}
  Mod.\ Phys.\ Lett.\  A {\bf 3}, 819 (1988).

\bibitem{David:1988hj}
  F.~David,
  {\it ``Conformal field theories coupled to 2D graviry
         in the conformal gauge,''}
  Mod.\ Phys.\ Lett.\  A {\bf 3}, 1651 (1988).

\bibitem{Distler:1988jt}
  J.~Distler and H.~Kawai,
  {\it ``Conformal field theory and 2d quantum gravity or
         who's afraid of Joseph Liouville?''}
  Nucl.\ Phys.\  B {\bf 321}, 509 (1989).

\bibitem{Furlan:2008}
P.~Furlan, V.B.~Petkova and M.~Stanishkov
{\it ``Non-critical string pentagon equations and their solutions,''}
Pre-print
{\tt [hep-th/0805.0134]}

\bibitem{BHKM:2007}
J.-E.~Bourgine, K.~Hosomichi,I.~Kostov and Y.~Matsuo
{\it ``Scattering of Long Folded Strings and Mixed Correlators in the Two-Matrix Model,''}
  Nucl.\ Phys.\  B {\bf 795}, 243 (2008).
{\tt [hep-th/0709.3912]}


\bibitem{Kharchev:2001rs}
  S.~Kharchev, D.~Lebedev and M.~Semenov-Tian-Shansky,
  {\it ``Unitary representations of $U_q(\mathfrak{sl}(2,\mathbb{R}))$,
         the modular double, and the multiparticle q-deformed Toda chains,''}
  Commun.\ Math.\ Phys.\  {\bf 225}, 573 (2002)
  {\tt [hep-th/0102180]}.
% SLE

\bibitem{Duplantier:2009}
B.~Duplantier, S.~Sheffield, {\it ``Duality and KPZ in Liouville Quantum Gravity,''} Phys.\ Rev.\ Lett.\ {\bf 102}, 150603 (2009) {\tt [0901.0277]}

\bibitem{Bauer:2008}
M.~Bauer, F.~David, {\it ``Another derivation of the geometrical KPZ relations,''} J.\ Stat.\ Mech.\ {\bf 0903} P03004 (2009) {\tt [0810.2858]}

\end{thebibliography}
\end{document}